\documentclass{aa}
\usepackage{natbib,psfig}  
\bibpunct{(}{)}{;}{a}{}{,}
\newcommand{\source}{{J1819+3845 }}
\begin{document}

\title{Annual modulation in the scattering of J1819+3845: peculiar plasma velocity and anisotropy}
\author{J. Dennett-Thorpe\inst{1,2}\thanks{jdennett@astro.uva.nl} 
\and  A.G. de Bruyn\inst{1,2}}
\institute{Netherlands Foundation for Research in Astronomy,
Postbus 2, 7990 AA, Dwingeloo, NL
\and Kapteyn Institute, Postbus 800, 9700 AV Groningen, NL
}

\date{Received ; accepted}
\titlerunning{Annual modulation in the scattering of J1819+3845}

\abstract{ We present two years of monitoring observations of the
  extremely variable quasar J1819+3845. We observe large yearly
  changes in the timescale of the variations (from $\sim$ 1 hour to
  $\sim$ 10 hours at 5\,GHz). This annual effect can only be explained
  if the variations are caused by a propagation effect, and thus
  affected by the Earth's relative speed through the projected
  intensity pattern. To account for this effect, the scattering plasma
  must have a transverse velocity with respect to the local standard
  of rest. The velocity calculated from these observations is in good
  agreement with that obtained from a two telescope delay experiment
  (Dennett-Thorpe \& de Bruyn 2001). We also show that either the
  source itself is elongated, or that the scattering plasma is
  anisotropic, with an axial ratio of $>$6:1. As the source is
  extended on scales relevant to the scattering phenomenon, it seems
  plausible that the anisotropy is due to the source itself, but this
  remains to be investigated. From the scintillation characteristics
  we find that the scattering material is a very strong, thin scatterer
  within $\sim$ten parsecs. We determine a source size at 5\,GHz of
  100 to 900\,microarcsecs, and associated brightness temperatures of
  $10^{10}$ to $10^{12}$K.
}

\maketitle



\section{Introduction}

Variability in radio sources can be due to either intrinsic variations
in the source itself, or due to a propagation effect known as
scintillation.  Disentangling these two effects is critical to our
understanding of the central regions of quasars and radio galaxies, as
in both cases the variability gives dimensions of the emitting region,
but the interpretation (and the implied size) are quite different.

Intrinsically variable extragalactic radio sources include blazars,
which exhibit changes in flux density of up to an order of magnitude
or more occurring over decades \citep{all85}, and smaller outbursts on
shorter timescales \cite[e.g.][]{poh96,ste95}. Intraday variability is a
phenomenon at GHz frequencies with typically a few percent variations
on day timescales \cite[e.g.][]{hee87,qui00}. An intrinsic explanation
requires the sources to have brightness temperatures far in excess of
10$^{12}$K -- up to $10^{17}-10^{19}$K.

Scintillation occurs due to changes in electron density in the
intervening material. Interplanetary scintillation is caused by the
solar wind. The variations of the flux density of quasars close to the
Sun was used to deduce that some quasars had components of arcsecond
size \citep{hew64,coh67,rea76} -- a claim since confirmed by radio
interferometry.  Interstellar scintillation is caused by the material
throughout the Galaxy, and is responsible both for many effects seen
in pulsars \cite[ and references therein]{ric90} and low frequency
variability (LFV) of some extragalactic radio sources
\citep{hun72}. The compact structure in LFVs is expected to be of the
order of a few milliarcsec, and is indeed seen in VLBI
observations\cite[e.g.][]{spa93}.  The scintillation regimes generally
appropriate for extragalactic sources are the (broad-band) weak and
refractive regimes. (Pulsar observations are generally concerned with
narrow-band diffractive phenomena. See Narayan, 1992 for a review of
scintillation).

Although refractive ISS is now commonly believed to be a major
contributor to the variability at low frequencies
\citep{spa89,spa93,bon96}, as first suggested by \citet{ric84}, its
relevance to observed variations at GHz frequencies has been the
subject of some debate \citep{wag95}, in particular because it cannot
account for the reported correlation between optical and radio
variability, which have been observed in a few sources
\citep{wag96,pen00}. The origin of the IDV in B\,0917+624, one of the
most variable classical IDV sources, with 10--15\% variations, has
been discussed by a variety of authors with different views
\cite[e.g][]{qui89b,qia91,sim91,sha91,ric95,kra99}. An intrinsic
explanation for the variability in this source gives an apparent
brightness temperature of 10$^{18}$K, which can be brought under the
Compton limit by a Doppler factor of $\sim$ 17 \citep{qui89b}.

PKS\,0405-385 was the first extragalactic radio source to show
variations at GHz frequencies that were fairly unambiguously
interpreted as due to ISS \citep{ked97}. Using the simplest assumptions
about the location and the velocity of the scatterer,
Kedziora-Chudzner et al. calculated the brightness temperature of
PKS\,0405-385 to be $5\times 10^{14}$K, or 1000 times the inverse Compton
limit. However, this figure depends critically on the assumptions
about the scatterer.

In this paper we investigate the properties of the flux density
variations of the quasar J1819+3845. The variations in this source are
the most extreme known in the radio sky. We previously interpreted the
variability as due to interstellar scintillation (ISS), based on the
extreme nature of the variations and their frequency dependence
\cite[ hereafter Paper I]{den00}. We have since proved them to be due
to ISS by detecting a difference in the arrival times (of up to
100\,sec) of the flux density variations at two widely spaced
telescopes \cite[ hereafter Paper II]{den01}. 

In Paper I we calculated the location of the scatterer using simple
formulae and the observed timescale, depth of the intensity variations
and the frequency of maximum variations. We assumed that the scatterer
was at rest w.r.t. the LSR, and thereby calculated in a straightforward
manner that the brightness temperature of the scintillating part of
J1819+3845 was $\ge 5\times 10^{12}$K. This is an order of magnitude
higher than the Compton limit-- a fact which could be explained by Doppler
boosting, but this expects rapid energy loss at higher radio
frequencies. Furthermore all known sources with Doppler factors of 10
or more are themselves intrinsically variable. 

We conducted a monitoring campaign on J1819+3845 to determine the
longevity and stability of the source and the velocity of the
scatterer (an unknown in the T$_B$ calculation). Since the programme
started we have measured the velocity of the scattering plasma by
another means (Paper II). With the monitoring campaign we also wished
to try to `resolve' the source by observing a change in scintillation
characteristics throughout the year, as we sampled the source along
different axes.  Here we report on the monitoring campaign at 5\,GHz
and will use the conclusions in later papers which will discuss the
(multi-frequency) structure of the source.

In Sect.~\ref{sec:obs} we describe the observations (and include an
optical spectrum). We give an overview of the results in
Sect.~\ref{sec:overview} before undertaking an analysis of the
light-curves (Sect.~\ref{sec:anal}).  Sect.~\ref{sec:vel} and
\ref{sec:aniso} give the interpretation of the results by fitting
models for plasma velocity and anisotropy to the data. In Sect. 7 and
8 give give some further consideration to the scattering medium and
source stability. In Sect. 9 we derive the parameters of the
scattering medium, and the size of the source, taking into
consideration effects of source resolution. We end with a summary of
the results and concluding remarks.

\section{Observations}
\label{sec:obs}
\subsection{Radio}

The observations were conducted with the Westerbork Synthesis Radio
Telescope (WSRT) in the period May 1999 to July 2001. The details of the
observations are given in Table~\ref{tab:obs1}. The telescope
generally observed in full array mode, with all antennae working at
the same frequency, but on a number of occasions two 
frequencies were observed separately in split-array mode. From
December 2000 onwards most observations were taken in split-array
configurations with 5 telescopes at 2.3\,GHz, 4 telescopes at 4.9\,GHz and
5 telescopes at 8.5\,GHz. The number of telescopes used at 4.9\,GHz
is indicated in Col. 4 of Table~\ref{tab:obs1}. (Data at the other
frequencies will be presented and analysed as part of a
forthcoming paper.) On-line integration times were set to 10\,seconds.
The total recorded bandwidth was 80\,MHz, in 8 contiguous
10 MHz bands. In the split-array observations software constraints 
limited the number of independent bands to 6$\times$10 MHz.  

Most observations consisted of long, uninterrupted tracks, of up to
12h duration. Each long track was preceded and followed by short
observations of polarized and unpolarized calibrators (generally
3C\,286/CTD\,93 before and 3C\,48/3C\,147 afterwards).  Absolute flux
density calibration was performed assuming a flux density of 7.42 Jy
for 3C\,286 and 1.70 Jy for CTD\,93 at a frequency of 4.874\,GHz.
The 
polarisation characteristics will be dealt with in a later paper.

The reduction package {\sc newstar} (Noordam et al., 1994) was used
to calibrate and examine the data. Bad bands, telescopes and baselines
were excised where necessary. \nocite{nor94}
 The amplitude and phase calibration
solutions determined for the calibrator were applied to J1819+3845.
The amplitude stability of the WSRT is generally very good. 
The errors generally contain three components which are due to 
pointing ($<$ 1\% at 4.9\,GHz), system temperature and atmospheric opacity 
variations.  
The main observed variations in T$_{\rm sys}$ are due to variable
spillover losses and atmospheric emission, both of which vary only
slowly. The WSRT continuously measures T$_{\rm sys}$ from stable noise
sources, allowing these slow variations to be accurately removed, resulting in an overall array amplitude stability of better than 1\% over
periods in excess of 12 hours.

The largest remaining uncertainty is due to the atmospheric opacity 
variations associated with (heavy) rainfall. The emission effects on the 
T$_{\rm sys}$ are taken care of but the associated opacity effects are
not. 
The uncertainty in the flux density scale due to opacity effects is
less than 1\%.  Errors associated with heavy rainfall, occurring less
than 1\% of the time, can, however, lead to an underestimate of the
flux density of a few percent.

Phase self-calibration was applied to the observations of \source,
with a 5 minute averaging interval. The source model of \source at
4.9\,GHz was a single point source at the phase centre, as there are
no sources detected above 1 mJy within the primary beam.  Previous
observations with the VLA had already confirmed that there is no
structure in the source on scales in between 0.3 and 3\arcsec, the
resolution of the WSRT at 4.9\, GHz.  The source flux density was then
calculated from the real part of the complex visibility, averaged over
the desired sampling interval (30 or 120\,s). This requires that the
phase is stable over the period of the self-calibration solution, a
condition seen to be held from inspection of the slowly changing phase
solutions. This method has the advantage that it does not require a
noise bias correction which is necessary if scalar averaging would be
used. 
Rare occasions of rapid atmospherically induced phase variations, were detected by comparison with the scalar averaged data: on 27/08/1999, 08/09/2000 and 13/10/2000, when the phase fluctuations result in a minor (a couple of percent) time-dependent variation in the calculated flux density.

Typically between 70 and 80 baselines contribute to the final
datum. With an rms noise of 20\,mJy per baseline, for 120\,s averaging 
times in one 10\,MHz band, we then achieved a typical noise in the 
source flux density of about 0.8\,mJy for eight 10 MHz bands and 120\,s
integration. This 0.8\,mJy is the thermal noise error. For the 
4 telescope (6 baseline), 60\,MHz observations from December 2000
onwards the error is about 4 times larger, i.e. 3 mJy. 
The final flux density error is the quadrature sum of a thermal error 
and the slowly varying gain error discussed above which we estimated
at  typically 1\%.  With a median flux density of the source of 200 mJy 
we can see that the overall error is never more than about 1.5\%. 
The smooth light curves, in periods of slow scintillations, show that our 
estimations of the errors are realistic.

For the 10\,minute observations conducted in September 1999, and in
two periods in September 2000 and November 2000, the errors in the
absolute flux density scale outweigh the errors from the fit.

\begin{table}
\caption{Observational parameters}
\label{tab:obs1}
\begin{tabular}{llllll}
\hline
\hline
Date      &day   & MJD  & duration& number & notes\\
          &number&      & (hrs)   & tels &\\
\hline
                                                                 
1999-05-13 & 133 & 51311  &  12   &   7  &Paper I\\ 
1999-05-14 & 134 & 51312  &  12   &   7  &Paper I\\    
1999-06-21 & 172 & 51350  &   2.5 &  14  &\\                              
1999-07-02 & 183 & 51361  &   2   &  14  &\\                              
1999-07-05 & 186 & 51364  &   5.5 &  14  &\\                              
1999-07-17 & 198 & 51376  &   6   &  14  &\\                              
1999-07-30 & 211 & 51389  &   6   &  14  &\\                              
1999-08-13 & 225 & 51403  &  12   &   7  &\\                              
1999-08-27 & 239 & 51417  &   6   &  14  &\\                              
1999-09-07 & 250 & 51428  &  12   &  14  &\\                              
1999-09-21 & 264 & 51442  &   5   &  14  &\\                              
1999-09-25 & 268 & 51446  &   5   &  14  &\\                              
1999-10-16 & 289 & 51467  &   4   &  14  &\\                              
1999-10-29 & 302 & 51480  &   7   &  14  &\\                              
1999-11-28 & 332 & 51510  &   6.5 &  14  &\\                              
1999-12-11 & 345 & 51523  &   8.5 &   7  &\\                              
1999-12-30 & 364 & 51542  &  12   &   7  &\\                              
2000-01-21 &  21 & 51564  &  12   &   7  &\\                              
2000-02-05 &  36 & 51579  &  12   &  14  &\\                              
2000-02-29 &  60 & 51603  &   2   &  14  &\\                              
2000-03-09 &  69 & 51612  &   2   &  14  &\\                              
2000-03-18 &  78 & 51621  &  12   &  14  &\\                              
2000-04-14 & 106 & 51648  &   9   &  14  &\\                              
2000-05-14 & 135 & 51678  &   6   &  14  &\\                              
2000-05-28 & 149 & 51692  &   8.5 &  14  &\\                              
2000-06-23 & 175 & 51718  &  12   &  14  &\\                              
2000-07-26 & 208 & 51751  &  12   &  14  &\\                              
2000-08-09 & 222 & 51765  &  12   &  14  &\\                              
2000-08-27 & 240 & 51783  &  12   &  14  &\\                              
2000-09-08 & 252 & 51795  &  12   &  14  &\\
2000-10-13 & 287 & 51830  &  10   &  14  &\\                              
2000-11-25 & 330 & 51873  &  12   &   7  &\\                              
2000-12-10 & 345 & 51888  &  12   &   4  &\\      
2000-12-26 & 361 & 51904  &  12   &   4  &\\                              
2001-01-07 & 007 & 51916  &   6   &  14  &Paper II \\        
2001-01-12 & 012 & 51921  &   6   &  14  &Paper II \\ 
2001-03-17 & 076 & 51985  &  12   &   4  &\\                              
2001-03-31 & 090 & 51999  &  12   &   4  &\\                              
2001-05-20 & 140 & 52049  &  12   &   4  &\\                              
2001-06-07 & 158 & 52067  &   9   &   4  &\\                              
2001-06-10 & 161 & 52070  &  12   &   4  &\\                              
2001-07-02 & 183 & 52092  &  12   &   4  &\\                              
2001-07-24 & 205 & 52114  &  12   &   4  &\\                              
\end{tabular}

\begin{tabular}{llll}
\hline
\hline
month & day numbers & duration & telescopes\\
\hline
1999-09 & 244,245,246,247,248,    & 10min &14\\
        & 252,254,256,261,262     &       &  \\
2000-09 & 244,245,246,247,248,249 & 3min  &14\\
2000-10 & 300,301,302,303,306     & 3min  &14\\
        & 307,308,309,311,312,313 &          \\
\hline  
\end{tabular}
\end{table}

\subsection{Optical}

\begin{figure*}[h!tp]
\psfig{file=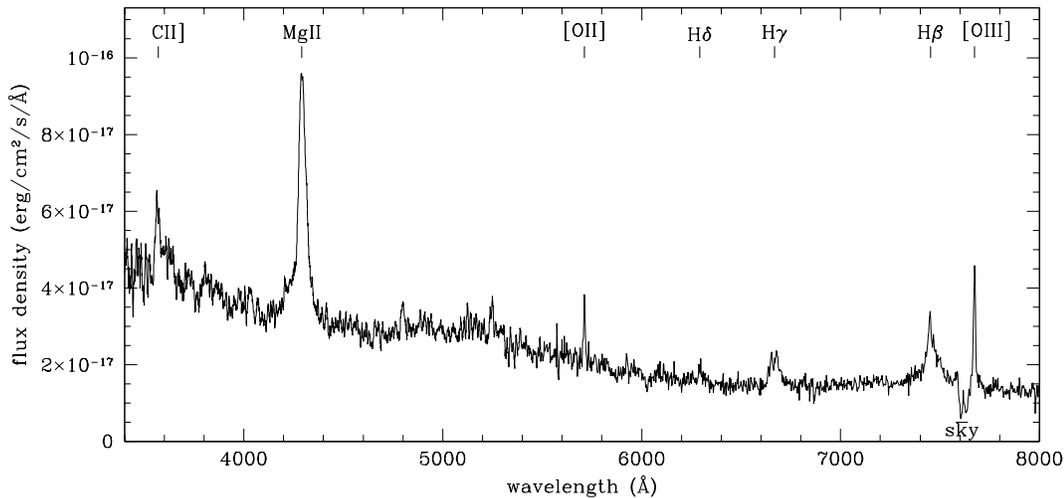,height=7cm,angle=0,clip=}
\caption{WHT spectrum of J1819+3845, with identified lines indicated}
\label{fig:spectrum}
\end{figure*}

A spectrum was obtained at the WHT on 17 July 2000 in three 1800\,sec
exposures. The ISIS spectrograph was used to obtain a spectrum from
3000\AA~ to 8000\AA, with a resolution of 0.86 and 2.9\,\AA/pixel and
central wavelengths of 4495\AA~ and 6708\AA~ on the blue and red arms
respectively. Standard reduction procedures were followed using the
{\sc iraf} {\em longslit} software package.  The combined spectrum,
with the blue arm boxcar smoothed over 9 pixels, is shown in
Fig~\ref{fig:spectrum}. The fitted line properties are shown in
Table~\ref{tab:opt}.

We calculate the redshift to be 0.533$\pm$0.001. The spectrum is
fairly typical for quasars, both in the equivalent widths of the lines
and in the continuum emission (see e.g. Miller et al. 1992; Baker
1997; Brotherton et al. 2001). We note the relatively small Balmer
decrement (H$\beta$/H$\gamma$), possibly indicating little line of
sight obscuration by dust. The [OII] luminosity (10$^{34}$W) is
comparable to moderately powerful FRII radio sources (radio
luminosities at 151\,MHz of $10^{25}-10^{27}$W/Hz/str)\citep{wil99}.
\nocite{bro01,mil92,bak97} We have observed the optical spectrum of
this source twice. On both ocassions the optical magnitude was
similar, and the lines strong and broad. From these observations it
does not seem that \source is a typical OVV; rather a typical quasar.

\begin{table}[h!tp]
\caption{Fitted line properties of optical spectrum}
\label{tab:opt}
\begin{tabular}{lllrrl}
\hline
\hline
lines & $\lambda$ & $\pm$ &EW$_{\rm rest}^{\rm(int)}$&EW$_{\rm rest}^{\rm(fit)}$& f \\
     &(\AA)  & (\AA)      &(\AA)& (\AA) &(erg/s/cm$^2$)\\
\hline                    
CII$]$    & 3565.5 & 1.0  & 8    & 7  & 3.7e-16\\
MgII      & 4292.8 & 0.3  & 73   & 69 & 2.4e-15\\
$[$OII$]$ & 5712.1 & 0.3  & 5    & 5  & 1.1e-16\\
H$\delta$ & 6293.2 & 2.0  & 5    & 3  & 0.8e-16 \\
H$\gamma$ & 6668.1 & 2.0  & 22   & 21 & 3.2e-16 \\
H$\beta$  & 7452.2 & 2.0  & 51   & 45 & 7.7e-16 \\
$[$OIII$]$& 7673.5 & 0.3  & 20   & 19 & 2.8e-16 \\
\hline
\end{tabular}
\end{table}


\section{Overview of radio observations and comments on individual epochs}
\label{sec:overview}

\begin{figure*}[h!tp]
\caption{Two years of monitoring observations of J1819+3845 at 5GHz.
  The first page covers 11 months, the second page 15 months. The annual change in the timescale of the modulations
  is clearly apparent.The horizontal axis shows the Hour Angle of the
source at the WSRT, so that each panel covers to 12 hours.}
\label{fig:res}
\psfig{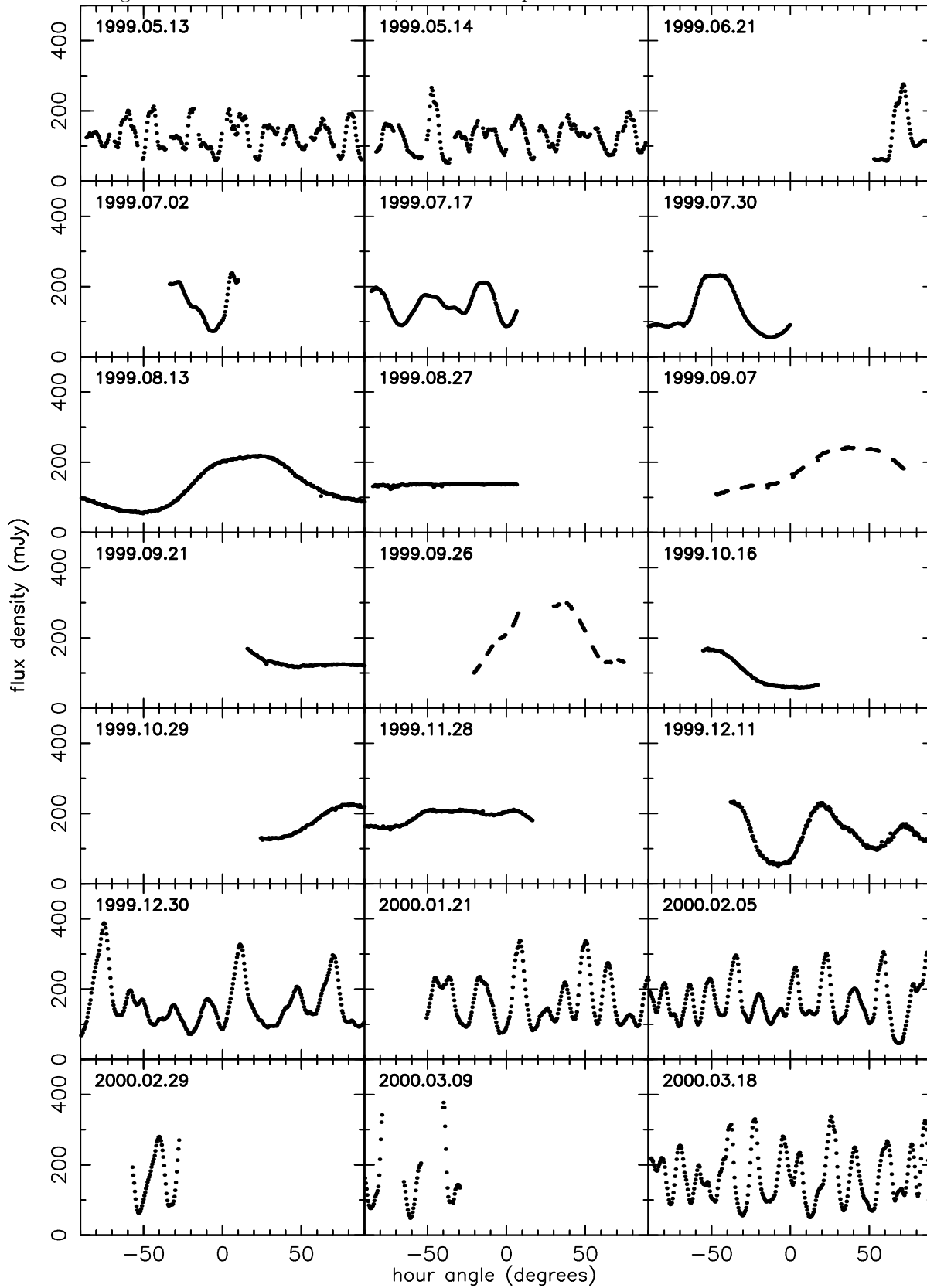}
\end{figure*}

In Fig.~\ref{fig:res} we present the light-curves over more than two
years of observations. In these curves the averaging time is
120\,s. For the analysis we used 30\,sec integrations, an example of
which is plotted in Fig.~\ref{fig:example}.  The interrupted
light curves in September 1999 are due to rapid switching between two
frequencies about 100-200 MHz apart.

\begin{figure*}[h!tp]
\psfig{file=h3795_f2b.ps,width=16cm,clip=}
\end{figure*}

The rapid variations of May 1999 have become slower by June and July
of that year, although they still are fast enough to show complete
modulations in a few hours. By the end of August the modulations have
become so slow that the source remains at an almost constant flux
density over 6\,hours. The fact that the source is still varying, but
with a very long timescale can be seen in Fig.~\ref{fig:res2}, which
shows the variations over August and September, including the short
($\sim$ 10\,mins) observations taken over the ten days following the
quasi-stationary observations on August 27 (day 239).  The source was
observed close to its minimum flux density (60\,mJy) on the next
observation (September 1, day 244), and on the following day at a
flux density typical of the peak excursions (220\,mJy).

The modulations remain slow until December, when they quickly speed up
again. By the end of 1999 there are rapid intensity variations
with peak to peak timescales of around one and a half hours.  The
variations remain fast, up to and including the observations of 15 May
2000, one year after the first twelve hour observations. It can be
seen that the character and timescale of the observations is extremely
similar to the observations of one year previously.

The observations during May 2000 to July 2001 broadly replicate the
observations of the previous year: showing a slowing down of the
variations between June and December, before speeding up again.  Note
also the very rapid speeding up of the scintillations in December of
both years: a factor of 2 within an interval of just 3 weeks.

 We have analyzed variations across the 80 MHz bands and found them
 highly correlated. The correlation coefficients of $>$ 0.99 are
 consistent with completely correlated signals with thermal noise. At
 5\,GHz the scintillation is clearly broad-band (Fig.~\ref{fig:eight}).

\begin{figure}[h!tp]
\psfig{file=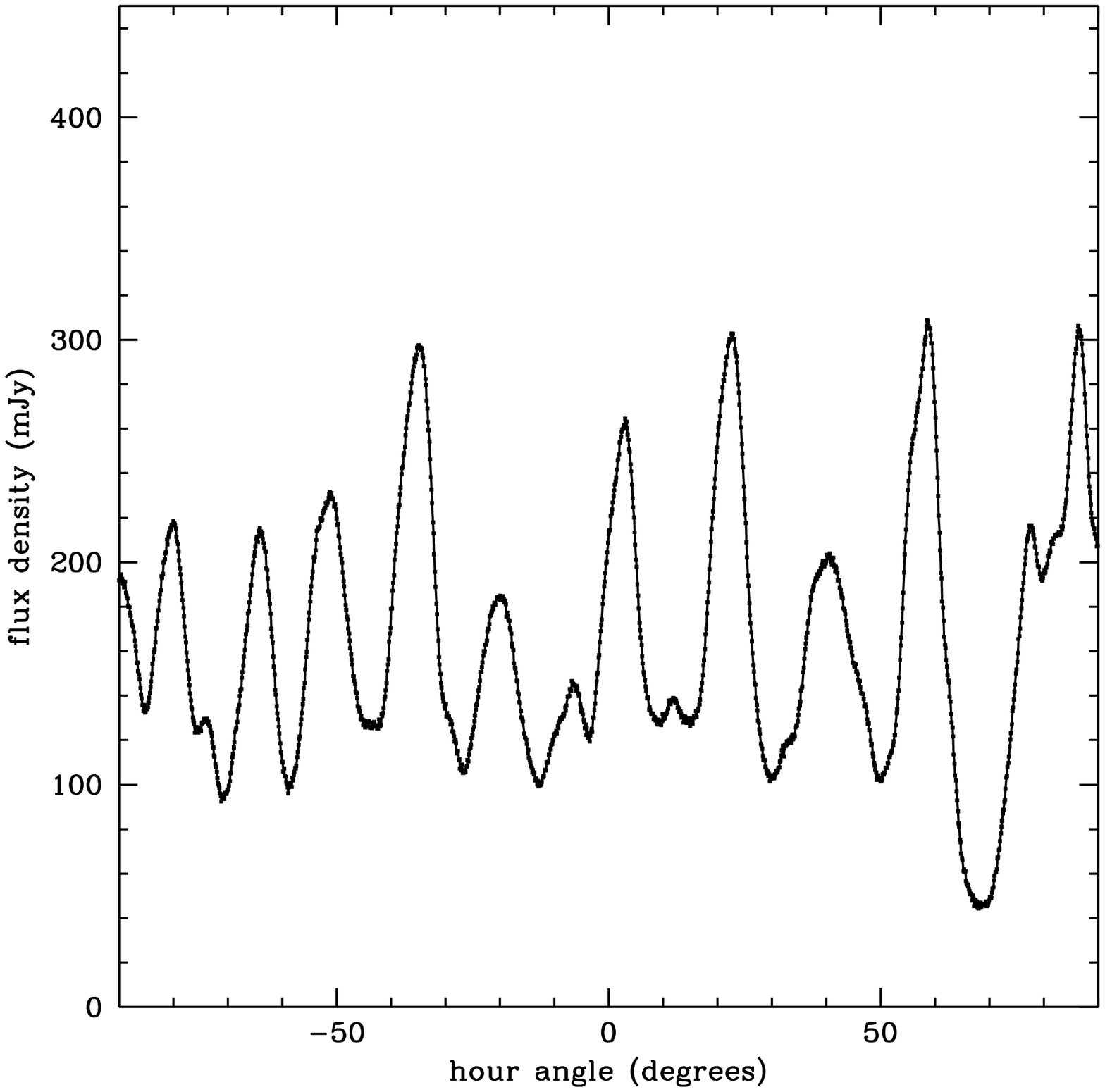,angle=0,height=8cm}
\caption{The light curve at 30\,sec resolution from 05 February 2000.}
\label{fig:example}
\end{figure}

\begin{figure}[h!tp]
\psfig{file=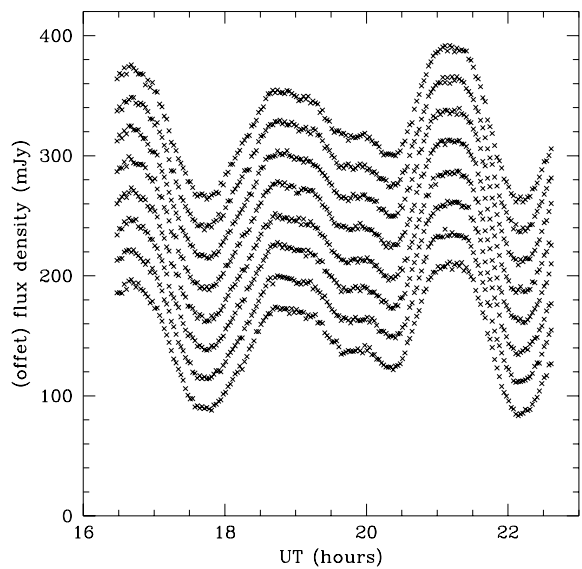,angle=0,height=8cm}
\caption{The light curves from 17 July 1999 at 120\,sec resolution showing all eight 10\,MHz bands.  The lowest light curve corresponds to a central frequency of 4.839\,GHz. The other bands, each 10\,MHz higher in frequency than the previous, are displayed with an offset interval of 25mJy.}
\label{fig:eight}
\end{figure}

\begin{figure}[h!tp]
\psfig{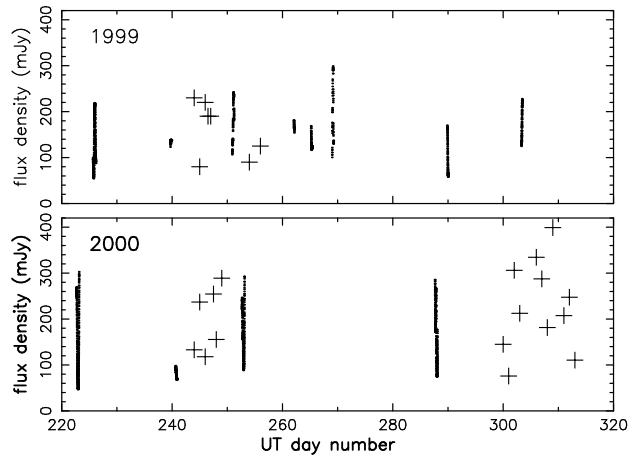}
\caption{Observations in August through October in 1999 (top panel)
and 2000 (bottom panel). In the 6-hour observation on 27 August 1999
(day 239), the source appears to maintain a near-steady flux density:
but the variations have not stopped, merely slowed down.}
\label{fig:res2}
\end{figure}


\section{Analysis of light curves}
\label{sec:anal}

\subsection{Modulation index}

The modulation index is the normalized rms intensity variations,
$\sqrt{\Sigma(\bar S - S)^2}/\bar S$. In the asymptotic theoretical
limit, the modulation index of broad-band variations of a point source
(a source subtending an angle much smaller than the scattering angle)
reaches unity at a `critical frequency'. Above and below this frequency
the modulation index drops.

We have calculated the modulation index for each of the observations
and present the results in Table~\ref{tab:res}. However, during the
period of slow modulations the `modulation index' calculated during
one run has little meaning, as typically only one scintle is seen per
observation. We therefore also present in Table~\ref{tab:res} a
modulation index calculated over the entire period July to October for
both years.

We calculate the error on the modulation index from Monte Carlo
simulations.  We used the observations which display over eight strong
intensity maxima, and selected portions of these runs, to
simulate the epochs where the variations are slower and we have few
scintles.
We calculate both an estimate of the scatter on calculated modulation
indices, and an estimate of the underestimation that is caused by
observing fewer scintles \citep{nar89,des90}. For an observation of N
high maxima (`scintles'), the scatter on the calculated modulation index is
given by $\sim0.3m/\sqrt{N}$, and the underestimation of the
modulation index is typically around half this. The error on the calculated mean flux density S is found to be $\sim S/6N$, or 2\%, whichever is the larger.

For the period of short observations we have calculated the mean flux
densities and modulation indices. These are
shown at the end of Table~\ref{tab:res}. As the observed flux density is
not normally distributed, we again estimate the error on the
calculated quantities by use of Monte Carlo simulations. This time
we drew a number of flux densities randomly from an observation with
many variations.
We conclude that the error on the calculated modulation index, $m$,
after N observations, can be expressed as $\sim 2m/3\sqrt{N}$. The
error on the calculated mean $\bar{S}$ can be expressed as $\sim
\bar{S}/(3\sqrt{N})$.

\begin{figure}[h!tp]
  \psfig{file=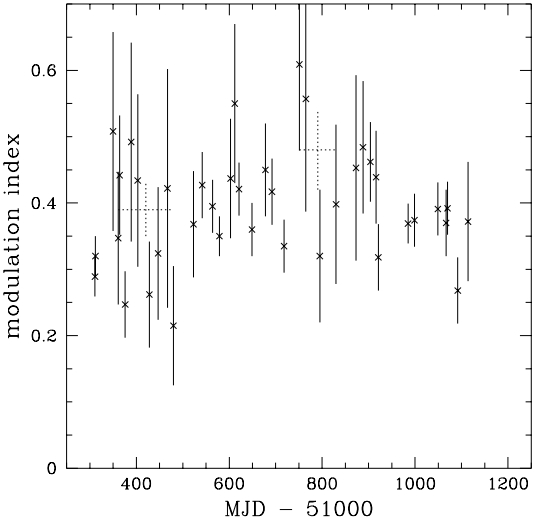,height=7.5cm}
\caption{Modulation index as a function of observing date. The mean
  flux densities used in this plot correspond to the mean measured
  during the respective observation. The two dotted crosses mark the
  modulation index calculated using all the observations in the slow
  period between July and October. Our estimation of the errors, given by the vertical bars, is explained in the text.}
\label{fig:mod}
\end{figure}

From this analysis we conclude that the modulation indices at 5\,GHz
are not statistically distinguishable throughout the observing period
(Fig.\ref{fig:mod}). They remain constant at $\approx$0.40.

\subsection{Timescales of variation}

The timescale of the variations can be defined in a number of
ways. The timescale used in Paper I was defined by use of the structure
function, as this is the quantity easily related to the theoretical
studies of scintillation. However this measure is impossible to
compute with less than one observed scintle. With very few observed
modulations any measure of the timescale will obviously be dominated
by statistical error. Therefore we have used independent ways of
estimating the timescale in an attempt to minimize the effect of this
on our conclusions. We describe these three ways of calculating the
timescale below, and present the results in Col. 8 to 11 of
Table~\ref{tab:res}.

\begin{figure}[h!tp]
\psfig{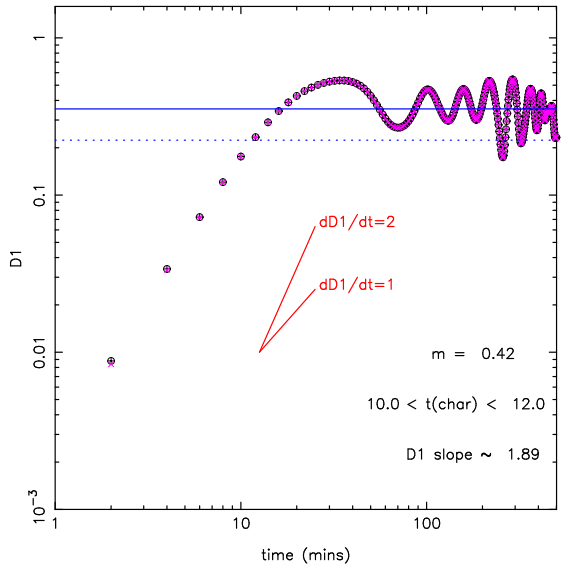}
\caption{A saturated structure function with observations from
18/03/2000. Crossed circles are points obtained directly from the data,
and the crosses are after correction for thermal noise (see text). The
error on the calculated timescale is dominated by statistical error
from a finite number of observed variations, not the thermal noise
correction. The solid lines indicate slopes $D^{1} \propto t^2$ and $t$ (see Sect. 7).}
\label{fig:sat-sf}
\end{figure}

 For our regularly sampled light curves f(i), i=1,2...n, the first
order structure function is estimated as
$$
D^1_f(k)=\frac{1}{\sum{w(i)w(i+k)}}\sum_{i=1}^{n}w(i)w(i+k)[f(i+k)-f(i)]^2$$
where w(i)=1 if the measurement exists and is otherwise zero
(Simonetti et al. 1985).  The structure function analysis is
straightforward for those observations with a large number of
modulations. In this case it can be seen that the structure function
has saturated (see e.g. Fig.~\ref{fig:sat-sf}), and we can calculate
the timescale as twice the time at the 1/$e$ of this maximum value
(which is D$^1$(max)= 2m$^2$). This is given by $t$ in Col. 8 of
Table~\ref{tab:res}.

Following the error analysis of
Simonetti et al. (1985) the observed value D$^1_f$(k) exceeds the true
value by an additional quantity 2$\sigma^2_{\delta f}$, where
$\sigma^2_{\delta f}$ is the measurement noise variance (assuming
white noise with correlation timescale smaller than the smallest time
interval measured). We thus assess the timescale correcting each value
of D$^1$ with the relevant noise variance, but the correction on the
calculated timescale is insignificantly small ($\sim$1\,sec for the
most rapid variations). The error on the calculated structure function
$D^1_f(k)$ from this contribution is negligible. From the Monte Carlo
simulations, we find that the (one sigma) error on this timescale due
to statistics of finite observing period is $\sim 0.25 t/\sqrt{N}$.

For observations with few ($\le$5) scintles we estimate a timescale,
from the structure function by assuming a mean modulation index for
all epochs. By considering the modulation indices obtained in May
1999, and December 1999 through May 2000, we take a modulation index
of $m$=0.4 for all epochs and estimate the timescale as the time at
which the structure function (or its extrapolation) reaches $2m^2$/e.
This modification is given by $t^\prime$ in Col. 9 of
Table~\ref{tab:res}.

As a final, and most direct, approach, we also estimated the
timescale, $t^\star$, from the average time between two minima (or
maxima) during the observation run. For runs with a small number of
scintles, $t^\star$ was calculated as the average of twice the time
between a minimum and maximum. We estimate errors on this process by
considering the runs with $>$5 complete modulations. We calculated the
$\sigma$ on the timescale as measured from each single complete
modulation independently. We found that the error on the measured
timescale can be estimated as $\sigma_{t^\star} = 0.3
t^\star/\surd{N}$, where N is the complete number of modulations
observed during the run. We calculated $t^\star$ in order to assess
the reliability of our error estimates on runs with few scintles, and
their possible propagation into the final results. The quantitative
results obtained using $t^\star$ are quoted in
Sect.~\ref{sec:aniso}, but we note here that the results so obtained
were similar to those obtained by the other estimations.  

For the observations of 9 March 2000, it is not clear how many minima
occur during the run, due to gaps in the observations. Considering the
rapid change of flux density on this date, it is plausible that extra
minima also occur during the unobserved periods. The timescale quoted
is based on this assumption, the error used in the fits is based on
the difference between this and the timescale calculated if only 4
minima occur during the observations. We can obtain lower limits to
$t^\star$ from the duration of the observations during which the
light curves showed little variation: 1999-08-27 (360\,min), 1999-09-21
(300\,min), 1999-11-28 (400\,min), 2000-08-27 (400\,min). This
information is not included in the fitting.

\begin{table*}
\caption{Derived quantities: Col. 5 is the number of scintles observed, N. Cols. 6 \& 7 list the modulation index and associated error. Cols. 8, 9, 11 show the derived timescales: using the structure function and derived modulation index, using m=0.4, and an estimation based on the average time between extrema, respectively.}
\label{tab:res}
\begin{tabular}{lrrrrrrrrrr}
\hline
\hline
Date & mean  & min & max &N &m& $\sigma_m$& t    & t$^\prime$ & $\sigma_{t}$ & ${t^\star}$\\
     & (mJy) &(mJy)&(mJy)&  & &          & (mins)& (mins)& (mins)     & (mins)        \\
\hline

1999-05-13 & 131.0 &  55.7 & 215.8 & 11.0 &  0.29 &   0.03 &   11.51 &   11.51&    0.87 &     76  \\
1999-05-14 & 126.6 &  47.3 & 271.1 & 10.0 &  0.32 &   0.03 &   14.71 &   14.71&
   1.16 &     77  \\
1999-06-21 & 129.5 &  55.5 & 277.3 &  1.0 &  0.51 &   0.15 &   18.85 &   14.00&
   4.71 &    110  \\
1999-07-02 & 152.7 &  70.4 & 239.4 &  1.0 &  0.35 &   0.10 &   26.00 &   32.00&
   6.50 &    156  \\
1999-07-05 & 144.5 &  58.1 & 277.0 &  2.0 &  0.44 &   0.09 &   29.43 &   27.30&
   5.20 &    180  \\
1999-07-17 & 148.6 &  83.3 & 213.5 &  2.0 &  0.25 &   0.05 &   26.24 &   55.00&
   4.64 &    270  \\
1999-07-30 & 125.9 &  53.8 & 234.9 &  1.0 &  0.49 &   0.15 &   51.59 &   41.50&  12.90 &    270  \\
1999-08-13 & 129.9 &  54.5 & 219.1 &  1.0 &  0.43 &   0.13 &  115.60 &  101.00&
  28.90 &    560  \\
1999-09-07 & 180.4 & 102.7 & 244.0 &  1.0 &  0.26 &   0.08 &  130.62 &  227.00&
  32.66 &    620  \\
1999-09-26 & 197.0 &  97.0 & 301.1 &  1.0 &  0.32 &   0.10 &   53.00 &   66.00&   13.25 &    400  \\
1999-10-16 &  97.9 &  56.1 & 172.1 &  0.5 &  0.42 &   0.18 &   82.12 &   77.00&   29.03 &    400  \\
1999-10-29 & 176.1 & 120.2 & 230.5 &  0.5 &  0.22 &   0.09 &   80.71 &  150.00&  28.54 &    550  \\
1999-12-11 & 139.8 &  45.5 & 247.0 &  2.0 &  0.37 &   0.08 &   42.29 &   46.90&    7.48 &    220  \\
1999-12-30 & 157.6 &  58.9 & 395.7 &  7.0 &  0.43 &   0.05 &   22.07 &   22.07&    2.09 &     97  \\
2000-01-21 & 168.3 &  69.7 & 342.7 &  8.0 &  0.40 &   0.04 &   14.78 &   14.78&    1.31 &     94  \\
2000-02-05 & 166.3 &  44.3 & 308.8 & 10.0 &  0.35 &   0.03 &   14.27 &   14.27&    1.13 &     68  \\
2000-02-29 & 157.8 &  61.0 & 281.6 &  2.0 &  0.44 &   0.09 &   12.76 &   11.20&    2.26 &     78  \\
2000-03-09 & 139.7 &  48.0 & 379.7 &  2.0 &  0.55 &   0.12 &    9.18 &    9.18&    1.62 &     51  \\
2000-03-18 & 168.5 &  48.7 & 342.1 & 12.0 &  0.42 &   0.04 &   11.65 &   11.65&    0.84 &     62  \\
2000-04-15 & 156.4 &  56.0 & 329.6 &  8.0 &  0.36 &   0.04 &   11.30 &   11.30&    1.00 &     68  \\
2000-05-14 & 143.3 &  42.5 & 314.2 &  4.0 &  0.45 &   0.07 &   15.42 &   13.06&    1.93 &    103  \\
2000-05-28 & 189.1 &  65.6 & 358.7 &  7.0 &  0.42 &   0.05 &   14.54 &   14.54&    1.37 &     78  \\
2000-06-23 & 209.4 &  64.4 & 403.4 &  7.0 &  0.34 &   0.04 &   20.67 &   20.67&    1.95 &     76  \\
2000-07-26 & 155.7 &  50.1 & 352.2 &  2.0 &  0.61 &   0.13 &   46.50 &   46.50&    8.22 &    256  \\
2000-08-09 & 147.4 &  44.8 & 303.3 &  1.0 &  0.56 &   0.17 &  127.79 &   87.00&   31.95 &    720  \\
2000-09-08 & 168.0 &  86.1 & 295.3 &  1.0 &  0.32 &   0.10 &  113.85 &  165.00&   28.46 &    720  \\
2000-10-13 & 163.2 &  72.2 & 287.1 &  1.0 &  0.40 &   0.12 &  124.00 &  124.00&   31.00 &    400  \\
2000-11-25 & 171.0 &  78.3 & 361.2 &  1.0 &  0.45 &   0.14 &  139.50 &  120.50&   34.88 &    370  \\
2000-12-10 & 190.1 &  32.6 & 376.3 &  2.0 &  0.48 &   0.10 &   68.72 &  68.72 &   12.15 &    290  \\  
2000-12-26 & 217.7 &  75.9 & 461.7 &  5.0 &  0.46 &   0.06 &   27.66 &  27.66 &    3.09 &    130  \\  
2001-01-07 & 179.3 &  45.3 & 362.8 &  4.0 &  0.44 &   0.07 &   21.98 &  19.00 &     2.75 &     75  \\  
2001-01-12 & 177.8 &  87.5 & 309.4 &  4.0 &  0.32 &   0.05 &   22.11 &  39.00 &    2.76 &    105  \\  
2001-03-17 & 228.3 &  66.4 & 527.6 & 12.0 &  0.37 &   0.03 &   11.62 &  11.62 &    0.84 &     50  \\  
2001-03-31 & 226.5 &  66.2 & 453.0 & 10.0 &  0.37 &   0.04 &   13.64 &  13.64 &    1.08 &     72  \\  
2001-05-20 & 223.1 &  69.9 & 503.6 &  8.0 &  0.39 &   0.04 &   15.54 &  15.54 &    1.37 &    110  \\  
2001-06-07 & 216.1 &  71.5 & 483.3 &  6.0 &  0.37 &   0.05 &   16.39 &  16.39 &    1.67 &     80  \\  
2001-06-10 & 207.7 &  79.4 & 507.7 &  8.0 &  0.39 &   0.04 &   17.45 &  17.45 &    1.54 &    120  \\  
2001-07-02 & 202.1 &  79.2 & 341.8 &  3.0 &  0.27 &   0.05 &   23.78 &  51.00 &    3.43 &    180  \\  
2001-07-24 & 183.2 &  95.0 & 394.5 &  1.5 &  0.37 &   0.09 &   52.33 &  59.00 &   10.68 &    480  \\  
\hline
\hline
\multicolumn{8}{l}{Characterization of short observations  (N = number of observations)}\\
1999.09.01 -- 09.19 & 159 &83.0 &234.0 &(10)& 0.32 & 0.07 \\
2000.08.31 -- 09.05 & 198 &118.0&289.0 &(6) & 0.36 & 0.10\\
2000.10.25 -- 11.07 & 228 &76.0 &399.0 &(11)& 0.43 & 0.08\\

\hline
\multicolumn{5}{l}{Combined data from slow season}\\
1999.07 -- 10  &148 & 53.8 &277.8  &  & 0.39 & 0.04\\
2000.07 -- 10  &158 & 44.8 &352.2  &  & 0.48 & 0.06\\
\hline
\end{tabular}
\end{table*}


\section{Detection of peculiar velocity of scattering plasma}
\label{sec:vel}

\begin{figure}[t!p]
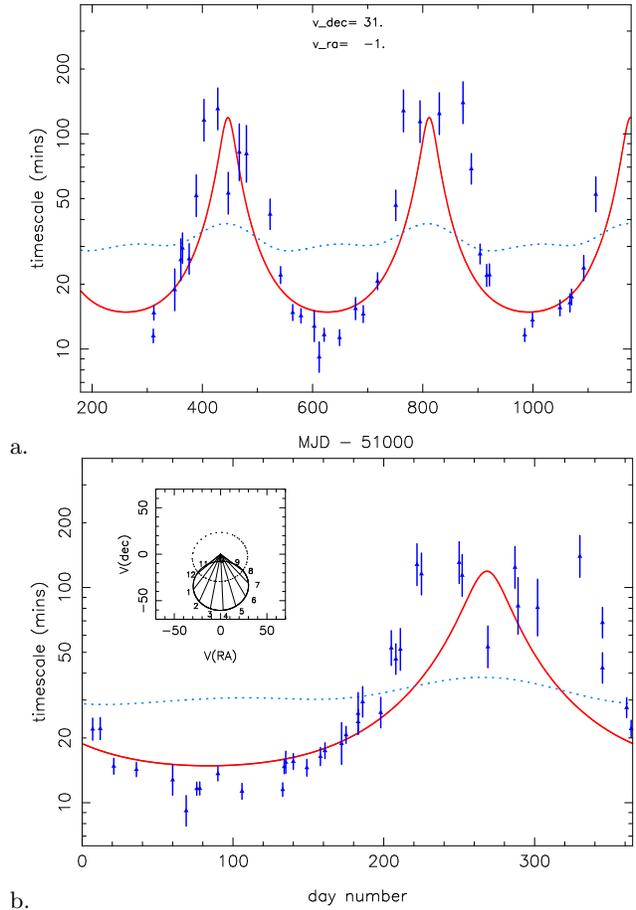

a.\psfig{file=h3795_f8a.ps,angle=-90,height=6cm}

b.\psfig{file=h3795_f8b.ps,angle=-90,height=6cm}
\caption{The timescales, t, (logarithmic scale) of the variations as a
function of observation date. The dotted line corresponds to the
prediction if the scattering plasma has no peculiar velocity.  The
solid line gives the best fit to the data (using timescale estimate t)
if the plasma is allowed a peculiar velocity.  The lower panel (b)
shows the data folded on a year. The annual modulation, as well as the
width of the epoch of slow modulations can be clearly seen. The
corresponding monthly velocity vectors $v_{\rm Earth} - v_{\rm
plasma}$ are shown in the ellipses: near-circular around the origin
for no plasma motion, and offset for the fit shown (see also Fig ~\ref{fig:briggs}).}
\label{fig:timescale}
\end{figure}

Fig.~\ref{fig:timescale} shows the changing timescale (t) as a function of
epoch throughout the period of observations. The yearly changes in the
timescale can also be clearly seen directly in the data (Fig.~\ref{fig:res}).

A change in the timescale of the scintillations with a year
periodicity must be due to the changing speed at which the Earth cuts
through the scintillation pattern projected onto the Solar System, as
we move around the Sun. The velocity of the scintillation pattern, as
viewed by a stationary observer on Earth, is given by the relative
velocity of the Earth and scattering plasma projected onto a screen
perpendicular to the direction of the source, $v_{\rm Earth} - v_{\rm
plasma}$.  We note that such a model predicts no change in the
modulation index. 

For most extragalactic sources, the vector $v_{\rm Earth}$ itself will
provide a substantial yearly modulation in the pattern speed.  This is
in contrast to pulsars, which produce scintillation patterns whose
observed motion is dominated by the pulsar's own velocity of typically several
hundreds km/s. One exception is PSR\,B2016+28, which has a transverse proper
motion velocity comparable to the Earth's.  \citet{gup94}
found a variation in the scintillation timescales of pulsars, which they
attributed to an annual variation caused by the Earth's motion. Unlike
most extragalactic sources, \source is {\it not} expected to show a large
annual modulation in the pattern speed. This is illustrated by the dotted
line in Fig.~\ref{fig:timescale}. \source is just 9$^\circ$ from the
direction of Sun's motion with respect to the LSR, and 28$^\circ$ from
the ecliptic pole, so the modulation due to the Earth's motion is slight. 

In order to fit the large change in timescale during the year we
require that the scattering plasma itself has a velocity with respect
to the solar rest frame (taken as moving at 19.7\,km/s towards 18 07
50, +30 00 52 (J2000)). The change in timescale throughout the year is
assumed to be entirely due to changes in our velocity through an
isotropic scintillation pattern. Thus we assume that the size of the
scintles, angular size of the source, and distance to the scattering
screen are constant throughout the years. However, we neither fix nor
determine these values, as the change in timescale only depends on the
changing projected velocity.

We probed a region of parameter space for the projected velocity of
the scattering plasma between -50 km/s and +50 km/s in both RA and
declination, and calculated the best fit. The best fits were clearly
constrained in this space, and the best weighted $\chi^2$ fit to the
calculated timescales is given by the solid line in
Fig.~\ref{fig:timescale}. It shows that the change in timescale can be
qualitatively explained by a velocity of the scattering plasma of
$\sim$ 30\,km\,s$^{-1}$. The model allows the timescale to be
normalised according to the best-fit. The ellipses in
Fig.~\ref{fig:timescale}b correspond to the projected velocities, for
the cases of no peculiar plasma velocity (dotted line) and a plasma
velocity corresponding to the fit shown (solid line). The velocity
vectors (using the fitted plasma velocity) are also shown for the
beginning of each month.

Although the model qualitatively describes the data, with an annual
period of long modulations, it is clear that the fit is poor (reduced
$\chi^2 = 9.1$ for $t$; goodness of fit parameter Q=10$^{-38}$).  The
fit is similar using the other timescale estimates.  The period of
very slow modulations extends for a much longer time than is allowed
by this simple model. Nor is the sharp reduction in timescale in
December, or sharp increase in timescale in July/August each year
replicated by this model. Furthermore the fitted plasma velocity does
not agree well with that measured from a delay experiment (Paper II).
However, a peculiar velocity of the scattering plasma is the only
possible explanation for this annual variation. In the next section we
consider further refinements and additions to the model which can
explain the long period over which slower modulations are observed.


\section{Detection of source structure/ anisotropic scattering medium}
\label{sec:aniso}

In the preceding section we assumed that the scintillation pattern was
isotropic. Anisotropic scintles could be produced by either an
extended anisotropic source or an anisotropic scattering medium. An
anisotropic scintillation pattern would leave a distinctive trace in
the form of an annual variation of the timescale.  This would happen
as the Earth moves around the Sun, and we cut the pattern at varying
angles through the anisotropic pattern, thereby introducing an annual
variation in the timescale of modulations.

\begin{figure*}[h!tp]
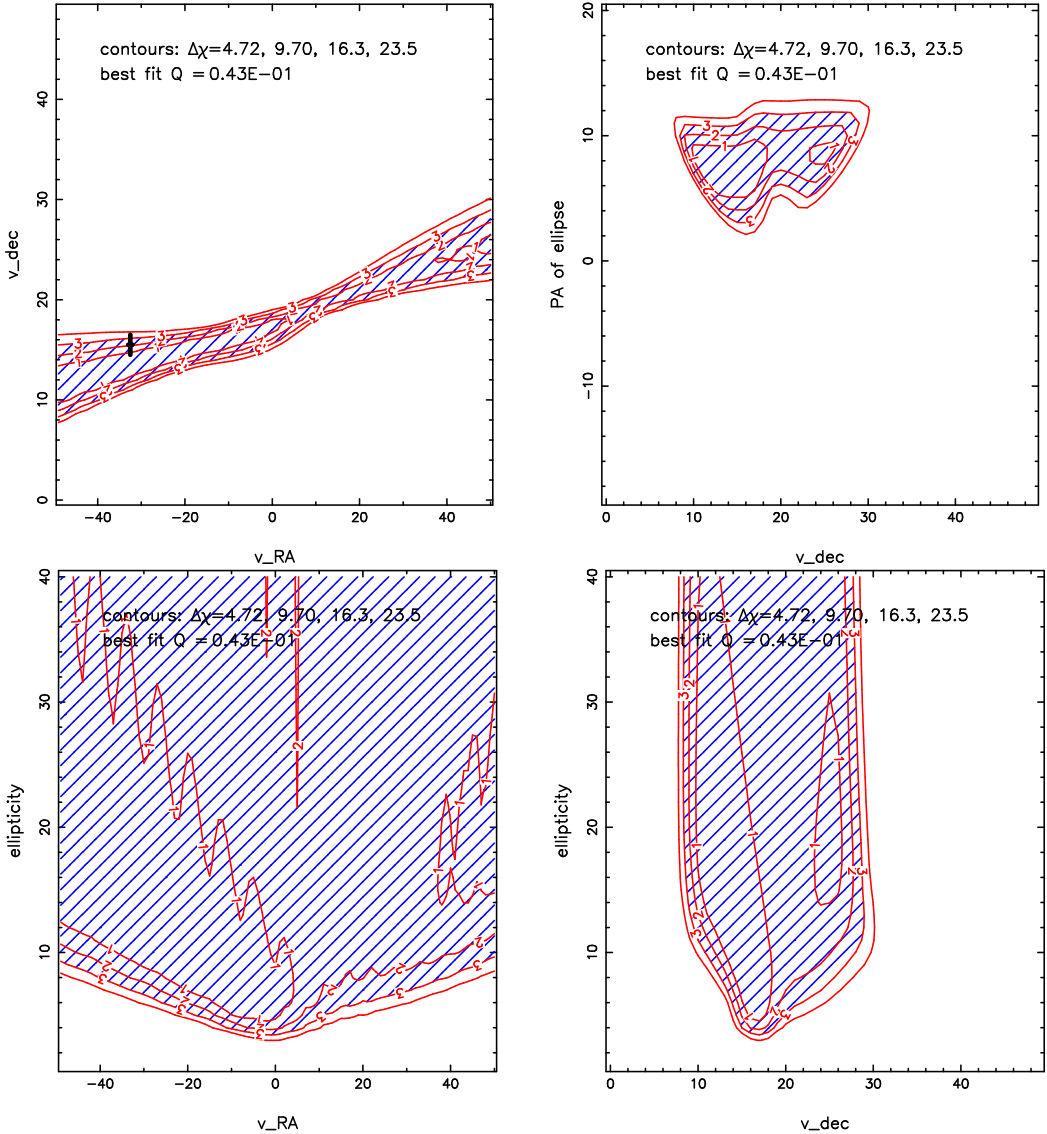

\psfig{file=h3795_f9a.ps,height=7.5cm,angle=-90,clip=}
\psfig{file=h3795_f9b.ps,height=7.5cm,angle=-90,clip=}
\caption{The probability distributions based on $\chi^2$-minimization
fits, using t. For each plot the probabilities are calculated over the
ranges shown in the other plot. The contours correspond to 68.3, 95.4,
99.73 and 99.99\% certainties for a normal distribution of errors.
The vertical line in the first panel represents the velocity and its errors derived in Paper
II. Note that in these plots PA is defined as E through N}
\label{fig:chi2}
\end{figure*}

In the absence of any peculiar velocity of the scattering plasma, the
Earth would cut through the scintillation pattern of J1819+3845 at the
same angle, but moving in different directions, approximately every
six months. Thus, for this source in the absence of any peculiar
scattering plasma, the lengthening of the timescale would occur about
every six months. It is therefore clear that source structure cannot
be the {\it only} explanation of the observed changes in timescale.

With the addition of a peculiar velocity of the scattering medium,
however, the periods of slow modulations no longer occur every six
months. It is therefore possible that both a peculiar plasma velocity
and source structure combine to produce the observed annual variation
in timescale.

We modelled an anisotropy in the scintillation pattern as an
ellipsoid, introducing two new parameters: the position angle and
axial ratio of the scintles. 
Using these extra parameters we again performed a $\chi^2$
minimization over a wide range of parameter space. The resulting best
fits are significantly improved over the fits without the anisotropy
term (reduced $\chi^2$ is 1.5 for $t$; 5.2 for $t^\star$; 7.1 for
$t^\prime$). Fig.~\ref{fig:chi2} shows the probability contours.
Although the fit was improved, good solutions were available over a
range of parameters.  It was found that the position angle and
v$_{\rm dec}$ were relatively well-constrained, but the axial ratio and
v$_{\rm RA}$ were poorly constrained.

\begin{figure*}[h!tp]
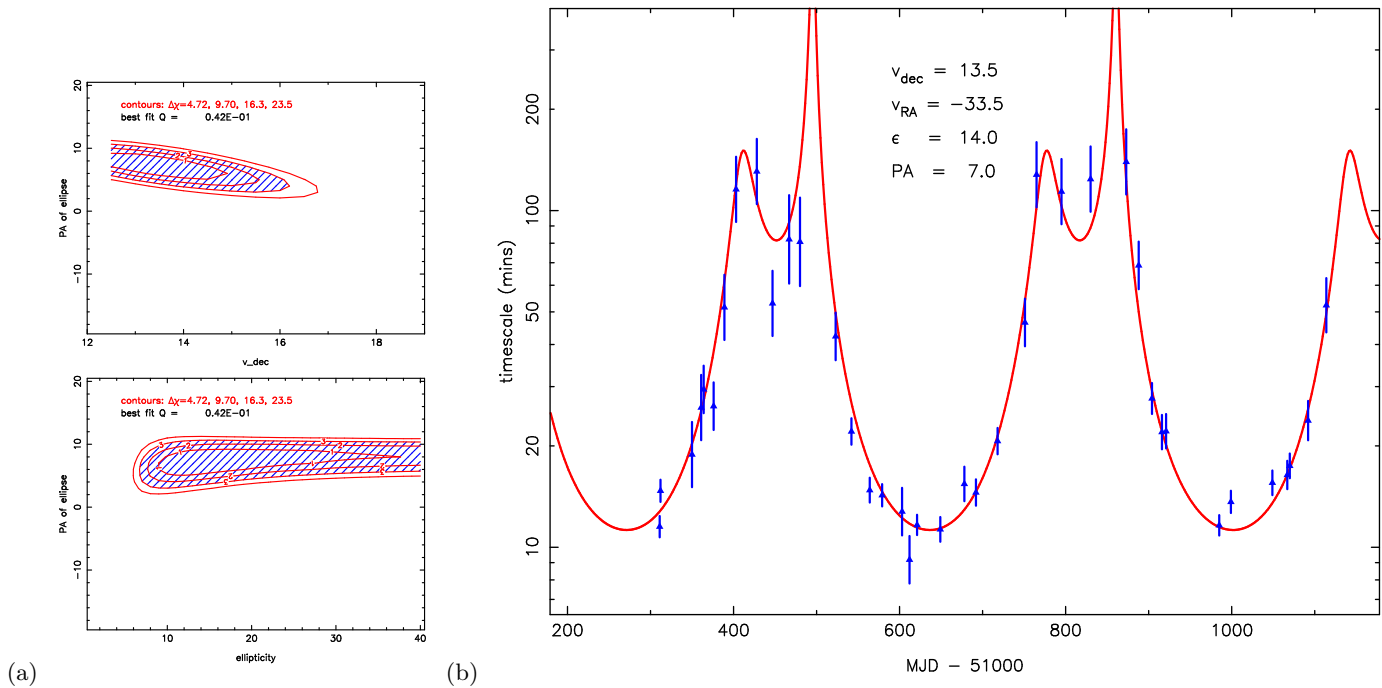

(a)
\psfig{file=h3795_f10a.ps,height=8.0cm,angle=-90,clip=}
(b)   
\psfig{file=h3795_f10b.ps,height=9cm,angle=-90,clip=}
\caption{Fitting constrained by the plasma velocity (and error) from
Paper II. Timescale estimate $t$ has been used.(a) The probability
distributions based on $\chi^2$ fits, showing the constraints placed
on the axial ratio and position angle of the anisotropy of the
scintles. (b) The resulting best fit to the timescales with epoch.}
\label{fig:fitfinal}
\end{figure*}

In Paper II we directly measured the velocity of the scattering plasma
using the time delay of signals arriving at two telescopes: v$_{\rm
  RA}=-32.5 \pm 0.5 $\,km/s; v$_{\rm dec}= +15.5 \pm 1$
km/s.\footnote{Note added in proof: We have since realised that full
  consideration of anisotropy in the two-telescope experiment is also
  needed. The effect of this is that the velocity is not as well
  constrained as given here, but the main conclusions are not
  significantly altered.} This result lies within the narrow region of
permissible v$_{\rm RA}$--v$_{\rm dec}$ space determined from these
yearly monitoring observations (see Fig~\ref{fig:chi2}).  Constraining
the velocity to be within the 3-$\sigma$ errors of this measurement,
we re-ran the fit and determined a probability profile for the shape
and position of the ellipse. This is shown in
Fig.~\ref{fig:fitfinal}a, and the results are listed in
Table~\ref{tab:fit}. The best fit (at v$_{\rm RA}=-33.5$ km/s; v$_{\rm
  dec}=+13.5$ km/s) has axial ratio 14$^{-8}_{(+>30)}$ at position
angle 83$^\circ \pm 4^\circ$ (N through E). Use of the timescales
obtained from the peak separation gives a similar result, although the
formal fit is very poor.

The introduction of two more parameters has clearly improved the fit
sufficiently to justify their use, as can be directly seen in
Fig.~\ref{fig:fitfinal}. We also note that the dates on which the
observations display very little variation over 6 hours, and which
therefore had no calculated timescale and were not included in the
fit, occur during the periods of predicted slow modulations. 

The fit
is illustrated in an illuminating way in Fig.~\ref{fig:squashed} which
shows the data compressed by the timescales predicted from the model,
such that the effects of plasma motion and anisotropy are removed.
After this timescale adjustment, the similarity of the modulations
throughout the two years can be clearly seen. The timescale of the
variations now looks similar throughout the period, indicating the
model fit was suitable. The nature of the variations is, by eye, also
similar throughout the observing period: there is no indication of big
changes in the `jaggedness', or the number and frequency of big
excursions. What is clear from the figure however, is the increasing flux
density of the brightest peaks in the light curves. The minimum flux
densities seem to have changed little, showing at most a slight
increase. This information is also reflected in Table \ref{tab:res}
[Cols. 3 \& 4]. As was shown in Fig.~\ref{fig:mod} there has been
no change of rms variations over time.  The change in appearance of
the light curves over time, in particular the increase in high maxima,
is due to the increasing flux density of the source (see
Sect.~\ref{sec:stab}).

One manner of producing the ellipsoidal scintles is by an ellipsoidal
source, as the scintillation intensity pattern produced by an extended
source is the convolution of the scintillation pattern of a point
source with the brightness distribution of the source itself. In the
strong, refractive regime a resolved source (source size $\theta_S >
\theta_{\rm scatt}$) is observed to have a timescale of modulation
larger than that of a point source as $\tau \propto
\theta_S/\theta_{\rm scatt}$ (Narayan, 1992), where $\theta_S$ is the
width of the source. If we assume that the anisotropy is then due to
an ellipsoidal source undergoing strong scattering, we may equate the
ellipsoid calculated above with that of the source itself. 

We note that we are observing near the critical frequency where the
asymptotic theory will not hold to great accuracy. However, as the
asymptotic theory from the other side of the critical frequency, in
the weak scattering regime, predicts $\tau \propto \theta_S^{7/6}$, we
take $\tau \propto \theta_S$ as a reasonable approximation. The second
complication is that to apply the theory as simply as we have done at
least requires the source to be larger than the scattering disk in all
directions. We believe however this is a reasonable assumption, as a
source that was significantly smaller than the scattering disk in one
direction should be expected to display different characteristics of
scintillations as the velocity vector cuts across its minor axis. At
this time it should display more of the second branch, or diffractive,
scintillations. There is no clear evidence for such a change in
character, such as a second, faster, timescale of modulation. Thus for
the anisotropy caused by an anisotropic source, in so far the choice
of ellipsoidal model is a good representation, the values obtained
(Table~\ref{tab:fit}) are expected to be close to the anisotropy of
the source itself.

An anisotropy in the scatterer could produce an effect similar to the
one described above for an anisotropic source.  Anisotropy in the
scattering medium is generally expected, and has been diagnosed by
elongated scatter-broadened images, with axial ratios of up to 3 found
through strongly scattering regions of the Galaxy
\cite[e.g.][]{lo93,fra94,wil94,tro98}, and axial ratios $>$ 5 in the
solar wind \cite[e.g.][]{nar89,arm90}. Compressible MHD turbulent
models have been developed in which the magnetic field gives rise to
an elongation of the electron-density fluctuations, with potentially
very large axial ratios \citep{lit01}. The effect of this anisotropic
medium on the scintillation observables is greatest when the screen is
thin, the magnetic field perpendicular to the line of sight and with
little change of magnetic field direction in the scatterer. In order
to produce the $>$ 6:1 axial ratio which would be required for
\source, we see from \citet{cha01} that the magnetic field in the
plane of the scattering screen must rotate through less than
$\pi/4$\,radians through the screen.  This implies that if anisotropy
in the plasma density structure causes the observed anisotropic
scintles, the magnetic field must be well ordered through the
scattering screen (i.e. along line of sight). The apparent $>$6:1
axial ratio required is somewhat unusual for high galactic latitudes,
but may be related to the unusual nature, and thinness of the screen.
\begin{table}
\caption{Results of fitting, using v$_{\rm RA} = -32.5 \pm 1.5$km/s;  v$_{\rm dec} = 15.5\pm 3$ km/s. Errors are 3-$\sigma$.}
\label{tab:fit}
\begin{tabular}{lllll}
\hline
\hline
           & axial ratio,$\epsilon$ & position angle &$\chi^2$ & Q \\
\hline
t          &14$^{-8}_{+(>30)}$    & $83\pm4$  &1.5 &4e-2\\
t$^\star$  &11$^{-4}_{+10}$      & $88\pm4$  &5.6 &2e-23 \\
t$^\prime$  &31$^{-10}_{+(>10)}$  & $83\pm4$  &7.2 &1e-32 \\
\hline
\end{tabular}
\end{table}

\begin{figure}[h!tp]
\psfig{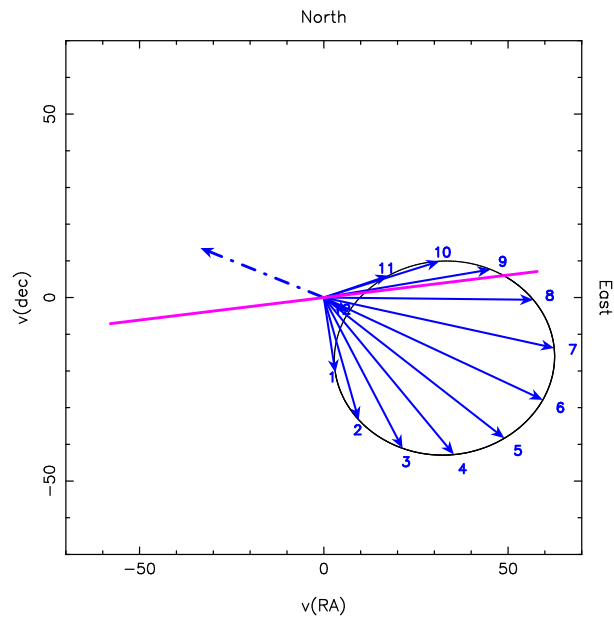}
\caption{The velocity vectors on the first of the month for the best
  fit model, using the constraints from Paper II. The velocity vectors
  are the velocity of the Earth through the scintillation pattern
  ($v_{\rm (Earth)}-v_{\rm (plasma)}$), as seen from the source.  The
  dashed-dotted line shows the projected plasma velocity ($v_{\rm
    (plasma)}$). In order to illustrate the combined effects of the
  velocity and the anisotropy we also present, on the same plot, the
  position angle of the anisotropy as seen from the source (solid
  line). }
\label{fig:briggs}
\end{figure}

\begin{figure*}
\psfig{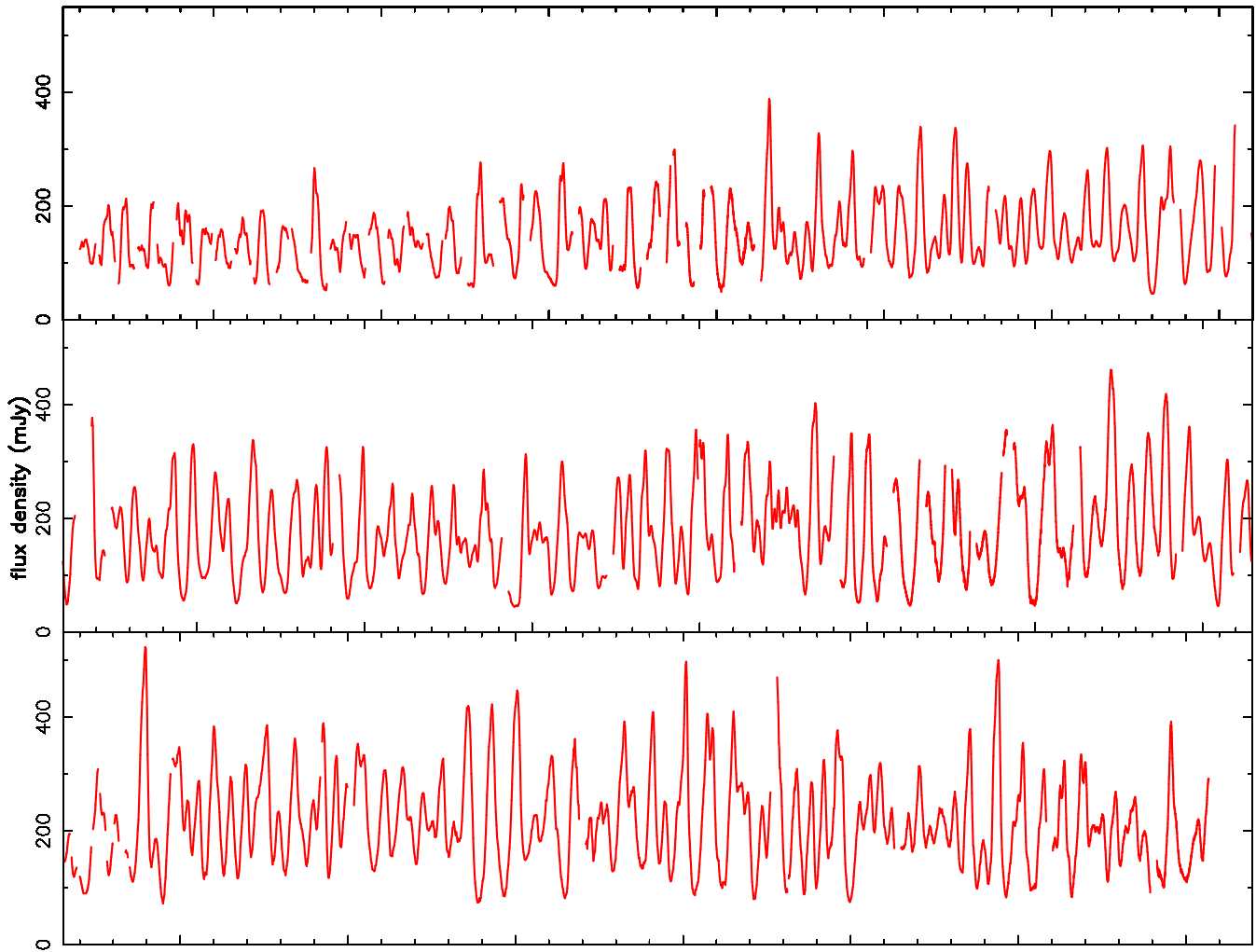}
\caption{The data adjusted by the timescales calculated from the best
 fit model.The timescale of the fluctuations is now constant. The
apparent change in depth of the modulations is an effect of an
increase in source brightness: the rms of the intensity fluctuations
about the mean does not change significantly over the years (see Fig.~\ref{fig:mod}).}
\label{fig:squashed}
\end{figure*}

\section{Grand Structure Functions}

In order to attempt to distinguish between the possibilities of source
or plasma anisotropy, to get clues about the depth of the scattering
plasma, we constructed grand structure functions $\langle
D^{1}(\tau)\rangle$.  For this purpose we took both the mean
calculated from the individual observations, and the mean obtained by a
linear fit to all the observations (Sect. \ref{sec:stab}). The data was
scaled in time according to the solutions from Sect. \ref{sec:aniso}. The
errors were calculated for each point directly from all contributing
data, weighted according to the number of points (after time scaling)
that contribute.

\begin{figure}
\psfig{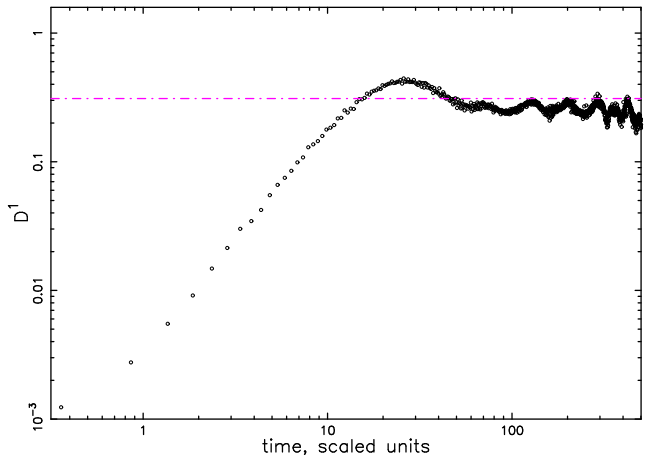}
\caption{Grand structure function, containing all the data, resampled according to the best fit from Sect. \ref{sec:aniso}. The dashed line corresponds to the saturation value calculated from the modulation index calculated from all (resampled) data.}
\label{fig:grand}
\end{figure}

Grand structure functions were calculated for all the data, and
over subsets of interest, including in and across the direction of the
elongation found above, and the individual years. We did not find any
very significant difference in the structure functions of these
sub-sets. We show in Fig.~\ref{fig:grand} the grand structure function
obtained using the entire data set.

The slope of the structure function before reaching saturation is
1.82$\pm$0.05 (i.e.$\langle D^{1}(\tau)\rangle \propto
\tau^{1.82}$).  This should be compared with ($\langle D^{1}(\tau)\rangle
\sim\propto \tau$) for classical IDV sources found by
\citet{sim85}. \citet{bla86} interpreted their result as due to
scattering in a thick screen, which they showed was able to reproduce
the $D^{1}\propto \tau$ result, unlike thin scattering screens which
give $D^{1} \propto \tau^2$.

On the basis of the slope of the structure function, it is therefore
likely that the scattering in J1819+3845 is predominantly in a
relatively thin screen (thickness $\ll$ distance). An inaccurate model
for the annual timescale variations would decrease the slope of
$\langle D \rangle$, but in the individual runs we did not observe
structure function slopes greater that 1.9.  We also note that the
grand structure function overshoots its saturation value. We consider
the origin of these features.

The $D\propto\tau^{1.82}$ relation could potentially be explained if
the screen, whilst thin, was allowed some finite thickness. We note
that $D\propto\tau^2$ for $\beta=4$, (where $\beta$ is the index of
the power spectrum of the density fluctuations) and shallower power
spectra (e.g. Kolmogorov) give rise to structure functions which rise
more rapidly. Therefore shallower power spectra cannot account for the
observations \cite[ Fig.~7]{rom86}. 

An anisotropic scatterer could produce a structure function which
overshoots its saturation value \citep{ric01a}. In this case we should
expect this to clearly show up as a difference in the structure
functions when moving across, and along the direction of
anisotropy. We see the overshoot when considering the entire dataset,
and no strong evidence for a difference between the structure
functions when sampling the scintles perpendicular and parallel to the
elongation. We conclude that preliminary analysis does not indicate
plasma anisotropy, however this result is tentative because of the
small number of scintles which contribute to the GSF parallel to the
elongation. Source elongation/structure and scattering plasma
anisotropy may be distinguishable by more sophisticated means (Rickett
et al, in prep).

\section{The stability and structure of the source}
\label{sec:stab}

\subsection{Source flux density changes}
\begin{figure}
(a)\psfig{file=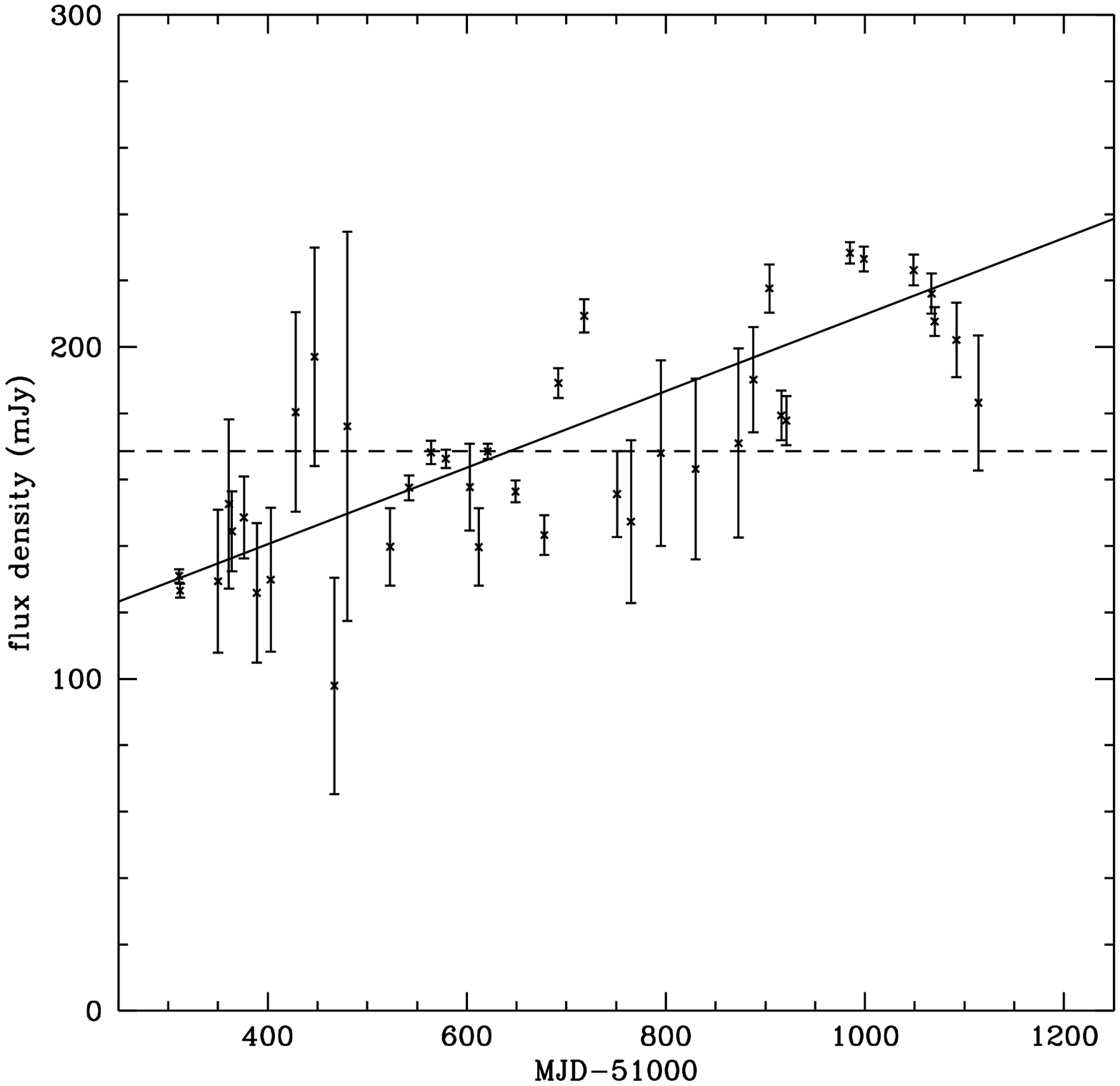,height=7.5cm}
(b)\psfig{file=h3795_f14b.ps,height=6cm,angle=-90}
\caption{(a)The mean flux density of source during the monitoring
  campaign. The solid line represents the best linear fit to the data;
  the dotted line the weighted mean.(b)The distribution of the
  observed flux densities in 1999, 2000 and 2001. To avoid
  over-representation of certain values, data from the slow periods
  has been sampled according to the timescales calculated previously.
  }
\label{fig:mean}
\end{figure}

The source has apparently undergone a change in mean flux density
during the two years of monitoring. Fig.~\ref{fig:mean} shows the
mean flux densities, with the error on them calculated as in Sect. 4.1.
Fitting a linear rate of change (using a weighted least squares fit)
gives a rate of change of intensity as $\sim$25\% per year. A line
with zero intensity change is a much poorer fit to the data.
The change in flux density is also clearly seen in the distribution of
observed flux densities (Fig.~\ref{fig:mean}b).  The mean flux
densities per year (data now sampled according to the timescale fit)
are 137\,mJy (1999$.$8), 172\,mJy (2000$.$5), and 215\,mJy (2001$.$3).

It is possible that the source became brighter at 5\,GHz in outbursts
at certain epochs: in particular around February 2001 (MJD $\sim$
51900). However, because of the extremity of the variations and the
number of scintles observed per run, we cannot distinguish this
possibility from a steady brightening.

\subsection{Source structure, and (lack of) changes}
\label{sec:resolv}

We have no evidence for structure on arcsecond or arcminute scales in
this source. From the radio spectrum (Paper I) we do not expect any
such structures to contribute more than a few mJy. We have looked for
evidence of extended structure by mapping several 12\,hour runs. We removed
\source by amplitude self-calibrating the whole run on the mean flux
density, and then looked for sources left in the field (which would
now themselves appear variable). There are a few faint compact sources more
than 1\,arcminute away, which have been seen also at 1.4\,GHz. We find
no evidence for any extended structure associated with \source to the
level of 0.1\,mJy for structure with scales 5--10\,arcsec.

From Sect. \ref{sec:aniso} we concluded that the scintle had an
elongated structure, but that this could be due to screen anisotropy
and/or source structure. If the anisotropy is due to source
structure it has been very stable, and by comparison of results from
two year periods, we conclude that it has changed position angle by
less than 10$^\circ$ in two years. 

Further information on the source structure can be gained from the
probability distribution function (PDF) of its flux density. A
point-like source has a very skew probability distribution function of
the intensities, and a very jagged appearance of the time series.
This is both expected in theory and seen in observations of pulsars
\cite[e.g.][]{ric00}. The smoothness of the time series -- the
quasi-oscillatory appearance -- indicates that the source is resolved
on the scale of a Fresnel zone at the scattering screen.

There is also an indication of asymmetry in the source brightness. The
falling edges of the flux density variations exhibit a more rapid
change than the rising edges \citep{deb01}. Initially a puzzling
result, because there was no evidence of a change in sense over the
year, this result can be explained with the scattering plasma
velocity. The scintillation pattern is always cut in the same sense,
and the Western end of the source is brighter than the Eastern. In a
future paper we will return in more detail to the scintle asymmetry.

We also use the PDF to look for changes in source structure over the
observation period.  We present the PDF for the entire data
(Fig.~\ref{fig:mean}b), where the data has been sampled according to
the theoretical timescale from the model fit in \S \ref{sec:aniso}.
The increase in brightness over the years is obvious. Normalizing on
the mean for each year (or the linear change calculated above), and
further increasing the sampling rate to produce independent points
(i.e. only one point in the calculated timescale is sampled) allows us
to calculate statistical measures of difference. Years 1999 and 2001
are statistically indistinguishable by the KS test ($>r$ 99\%
certainty).  However year 2000 has a 20\% probability (KS test) that
the distribution is similar to the other two purely by chance.
This is primarily due to the higher dispersion about the mean ($\sigma
= 0.38, 0.44, 0.39$ in 1999, 2000, \& 2001 respectively).  From this
preliminary investigation we see no evidence for change in structure.

In Paper I we considered the source to be composed of a scintillating
component and a non-scintillating base component. However, the source
may be a few times larger than the Fresnel scale, and all
scintillating. This would also result in modulations in flux density
that do not drop to zero, giving an appearance of a `base level'. As
previously noted, the source is expected to be resolved in order to
explain the smoothness of the variations. The 50\% increase in source
brightness with no concomitant change in modulation index, and indeed
the lack of difference in the normalized flux density PDFs over the
years, is difficult to explain in the `baselevel plus small component'
model. It would require that both the non-scintillating and the
scintillating components increased in flux density by a similar
amount, which is somewhat contrived, whereas this situation is a
natural result of a resolved source that gets brighter.  For this
reason we now favour the picture of a resolved source, a few times the
Fresnel size.

We then can interpret the modulation index in order to obtain the
source size relative to the scattering zone, which we will return to in Sect. \ref{sec:interp}.

\subsection{Expansion rate}

From the modulation index alone (Fig.~\ref{fig:mod}), which shows no
signs of changes over the two years, we can put an upper limit on the
source expansion. As we have reason to believe that the source is not
much smaller than the scattering disk (above), we can directly
translate the lack of decrease in modulation index to a lack of
detectable source expansion.  Because the scintillation pattern is a
convolution of the scintillation pattern produced by a point source
and the source brightness distribution, an increase in source size
would always act to decrease the observed modulations. In the strong
scattering regime, the modulation index $\propto$ 1/(source size)
\citep{nar92}. We put a limit on the decrease in modulation index at
10\%, and thereby calculate that the source has expanded by less than
10\% over two years.

\section{Properties of the scatterer and source size}
\label{sec:interp}

In Paper I we used a simple formula based on the observed timescale,
the frequency at which the modulations were maximum ($\sim$5\,GHz),
and an assumed velocity through the scintillation pattern to deduce the
location and strength of the scattering screen and the size of the
source. (note a typographic error in Paper I: C$_N^2$ calculated in
that paper should read 0.017, not 0.17\,m$^{-20/3}$). We will return to
this calculation again, now having ascertained the velocity through
the scintillation pattern as a function of epoch. 

As the velocity of the plasma screen is of the order a few tens of
km/s, and because of a lucky accident of the date of our first
observations, the velocity used in Paper I was a reasonable figure,
and therefore the measured velocity of the plasma screen does not
itself change significantly the apparent size of the source. However,
we also need to take in account the effects of the source size and the
anisotropy in the scintillation pattern. These introduce a number of
significant changes.

We reformulate the scattering results for a point source from Paper I,
in terms of the desired quantities. These formulae are derived from
Blandford \& Narayan (1985) and Romani et al.
(1986),\nocite{bla85,rom86} using the point of maximum refractive
scattering (m=1), and assuming that the timescale is just the time
taken to move through a scintle.
$$\theta_{\mu as}=64\,\lambda^\star_{cm}\,v^{-1}_{kms^{-1}}\,t^{-1}_{hrs}$$
$$C_{-4}=1.2\,10^8\,{\lambda^\star_{cm}}^{-1}\,v^{-11/3}_{kms^{-1}}\,t^{-11/3}_{hrs}$$
$$L_{kpc}=1.9\,10^{-4}\,v^{2}_{kms^{-1}}\,t^{2}_{hrs}\,\lambda^\star_{cm}$$ Here
C$_N^2$=10$^4$\,C$_{-4}$, $\theta$ is the radius of the scattering
disk, or the Fresnel disk at the critical frequency ($\nu^\star =
c/\lambda^\star$), and the distance to the equivalent screen is $L$,
and we move with velocity $v$ through the scintillation pattern with
timescale $t$. These equations can only be applied directly to
observations at the critical frequency for a point source, or to a
source with a non-scintillating base component and a point-like
scintillating component.

As argued in Sect. \ref{sec:resolv} however, we consider the source
to be resolved.  The introduction of a source size means that the
observables ($m$, $t$, $\nu^{\rm o}$) no longer reflect those of the
refractive scintillations from a point source. We cannot therefore just drop the brightness
temperature from that calculated in Paper I by a factor 4 to
compensate for a source size increase of a factor 2. In what follows
we will turn the observed quantities into the equivalent point source
values for use in the above equation, using the suffices $^\star$ and
$^{\rm o}$ to refer to quantities measured at the frequencies of maximum
modulations for a point source, and for an extended source
respectively. We refer to the frequency of maximum modulations for a
point source, $\nu^\star$, as the `critical frequency'.

\subsection{Effects of source size on $m$ and $t$ at maximum modulation}

\begin{figure}
\psfig{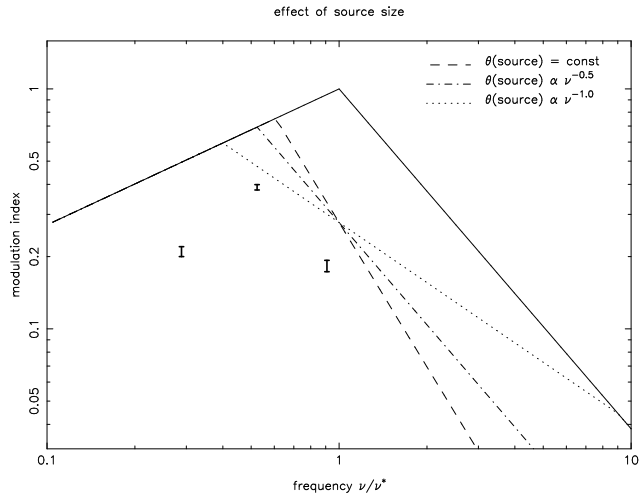}
\caption{The change of modulation index with frequency for source
sizes larger than the scattering disk at the critical frequency,
$\nu^\star$. The solid line represents a point source, the dotted
lines extended sources, with frequency dependencies as indicated. All
extended sources are 3 times the Fresnel zone at the critical
frequency ($\theta_F^\star$). The points correspond to multifrequency
observations, placed on the graph such that the 5GHz point is at
$\nu^{\rm o}$ for the case $\theta \propto \nu^{-0.5}$.}

\label{fig:meffect}
\end{figure}

A physical source would be expected to change its extent with
observing frequency. This has an effect on the characteristics of the
depth of modulation as a function of frequency if the source is
resolved ($\theta_S > \theta_{\rm scatt}$). An extended
source does not exhibit maximum intensity variations at the same
frequency as a point source, but at a frequency which is also dependent
on its change of size with frequency. This is because the observed
frequency of maximum modulations for a point source occurs when the
scattering disk is the same size as the Fresnel zone, but for an
extended source this occurs when the scattering disk is equal to the
source size. This is illustrated in Fig.~\ref{fig:meffect}

We are observing close to the frequency of maximum modulations.
Simultaneous multifrequency observations from 10/2000 through 07/2001
confirm the effect reported in Paper I. The observations will be
reported in a later paper, but the average modulation indices are
plotted on Fig.~\ref{fig:meffect}, as it is critical to our analysis
in this paper that the 5\,GHz observations are close to the modulation
maximum. We can calculate the frequency of maximum modulations that
would be observed for a point source, if we assume that the source is
larger than the scattering disk in the observed frequency regime, and
that there is no non-scintillating component.

We use the asymptotic theory \citep{nar92}, for both the strong and
the weak regimes. Calculating the predicted modulation indices for a
resolved source, we find that for source size decreasing with
frequency less rapidly than $\nu^{-12/7}$, the observed frequency of maximum modulations is 
$$\nu^{\rm o} = \nu^\star (m^{\rm o})^{30/17}$$
For illustration, the effect on
the modulation index as a function of frequency for a number of source
size frequency dependencies is illustrated in Fig.~\ref{fig:meffect}.
The frequency dependence of the source size does not, however, affect
the above relation.

Note that a non-scintillating component of the source would reduce the
observed modulations without changing the observed modulation maximum
from the critical frequency, but {\it a resolved source results in a
  true critical frequency which is above the observed frequency of
  maximum modulation}. This is because in this case, the maximum
modulations do not occur at the transition from strong to weak
scattering, but when the source size equals that of the scattering
disk. Based on this simple calculation which assume the asymptotic
theories are reasonable approximations to reality, and assuming there
is no non-scintillating component, and using the observed maximum in
the intensity modulations ($\nu^{\rm o}$) is around 5\,GHz, we find that the
critical frequency ($\nu^\star$) is around 25\,GHz.

We next consider the effect of source resolution on the observed timescale.
Using the relation between the observed frequency of
maximum modulation and the critical frequency derived above, we find
that the timescale observed at the maximum modulations is related to the timescale for a point source at the critical frequency, $t^\star$, by
$$t^{\rm o} = (m^{\rm o})^{-66/17}t^\star$$
Under the same assumptions as above, and using the
observed $m^{\rm o}$=0.4 we find $t^\star = t^{\rm o}/35$.

In order to relate the source size at the critical frequency to that
at the frequency of observed modulation maximum, we must assume a
dependency of source size on frequency.  From multifreqency monitoring
over the same period (forthcoming paper) we find that the spectral
indices of the source around 5\,GHz are $\alpha^{4.9}_{2.3} \approx
+1.0$ and $\alpha^{8.5}_{4.9} \approx +0.6$.  We therefore expect that
the source will display a frequency dependence between that of an
optically thick source ($\alpha=+2.5$, $\theta$ constant) and a
standard equipartition flat-spectrum source ($\alpha = 0$, $\theta
\propto \nu^{-1}$).  We therefore consider a source which decreases in
size as $\nu^{-0.5}$ (See also e.g. de Bruyn (1976)).\nocite{bru76}

In this case, and again for no non-scintillating component in the
source, we find that the maximum modulation index, $m^{\rm o}$, is related
to the source size and the Fresnel scale at the critical 
frequency as $$m^{\rm o} = (\theta_F^\star/\theta_S^\star)^{1/3}$$  Using the
observed $m^{\rm o}$=0.4, we find $\theta_S^\star = 15\theta_F^\star$. Thus
the source could be up to 15 times the size of the Fresnel scale, at
the critical frequency, i.e. at 25\,GHz.

\subsection{Effects of anisotropy}

Finally we must account for the anisotropy. Without consideration of
the anisotropy we cannot turn the observed values into source and
screen parameters, because of the order of magnitude change in
observed timescale during a year. The interpretation of the anisotropy
is the most difficult, and potentially the biggest source of error as
the theory used is an isotropic scattering theory. In what follows we
simply take geometric means of relevant quantities, and assume these
would be the values appropriate to an isotropic theory.  We consider
two cases: that the anisotropy is due entirely to the medium, or
entirely due to the source.  It is possible that both the effects play
a role: in this case the position angle of the anisotropy fitted to
the data represents a position angle of the convolution of the two
effects. 

If the anisotropy is due to the scattering medium, it is not clear how
we should take this into account, as a full treatment of anisotropic
scattering should be used, not the isotropic case on which we base our
arguments. It is complicated by the fact that, as argued above, the
source is likely to have a size comparable to the scattering scales.
To make a step towards accounting for the anisotropy, we consider the
situation when the Earth cuts through the anisotropy such that the
scale is the geometric mean of the two scales (day 220, for best fit
$\epsilon$= 14), when our velocity through the scintillation pattern
is calculated to be 54\,km/s.

\subsection{Derived source and screen parameters}

First we consider that the medium is isotropic and the source is a
resolved ellipsoid with no non-scintillating base component. Because
of the conversion estimates we are making to treat the isotropic
theory in a way to account for anisotropy, the results will also be
those for the case in which the observed anisotropy is due to the
medium, but the source itself is a partially resolved isotropic
source.

From the above considerations of the effect of source size we find
$t^\star= t^{\rm o}/35=0.057$\,hrs.  We note that here we associate twice
the timescales measured and presented in Table \ref{tab:res} with the
time required in the formula, and the angular scales as corresponding
to the diameter of the source. From the preceeding discussion, we
associate $t$ with $t^\star$ and $\theta$ with $\theta^\star_F$ in the
formulae given at the beginning of Sect. 9. Using $v$=54\,km/s,
$m^{\rm o}$=0.4, $\nu^{\rm o}$=5\,GHz in these formulae, we find $\theta_F^\star$=
25$\mu$as.  At 5\,GHz the source is then $\theta_F^\star \times 15
\times (\nu^\star/\nu)^{0.5} =860\mu$as.  The source has a mean flux
density of 200\,mJy, resulting in a brightness temperature $T_B= 2
S\lambda^2/\pi k\theta_S^2 =1.3\times10^{10}$ K. In this case the
screen is extremely strong (C$_N^2 \approx 165 m^{-20/3}$), and nearby
($\sim$ 1\,pc), which results in the larger angle subtended by the
scattering disks.  This in turn considerably lowers the brightness
temperature from that calculated in Paper I.

We now consider the case of the maximum possible flux density in a
non-scintillating component. Although the modulation index is 0.4, we
know that $>$40\% of the source is scintillating, by the regular
minima at $\sim$ 60\,mJy. We therefore consider, in a similar manner
to Paper I a non-scintillating base-level determined by this value.
From Fig.~\ref{fig:mean}(a) we take 140\,mJy in the scintillating
component. However our approach here differs from Paper I, not just
due to our revised velocity, and calculated anisotropy, but because we
take into account that this scintillating component must also be
somewhat resolved. The reason for this is that the modulation index is
not 0.7, as would be expected from scintillation of a 140\,mJy point
source at the critical frequency, on a non-scintillating component of
60\,mJy. The observed modulation index of the smaller component, i.e.
if we remove the base component, is ${m^{\rm o}}^\prime=m^{\rm o}/x$, where $x$ is
the fraction of the total flux density in the scintillating component.
We use ${m^{\rm o}}^\prime$ in place of $m^{\rm o}$ in the formulae given in the
preceding sections.  Again using v=54\,km/s, $t^{\rm o}$=2\,hrs, and
$\nu^{\rm o}$=5\,GHz, we obtain at 5\,GHz a source size of 100$\mu$as and
brightness temperature of $7\times10^{11}$ K. The scattering screen is
12\,pc away with C$_N^2=0.5$m$^{-20/3}$.

\begin{figure}
\psfig{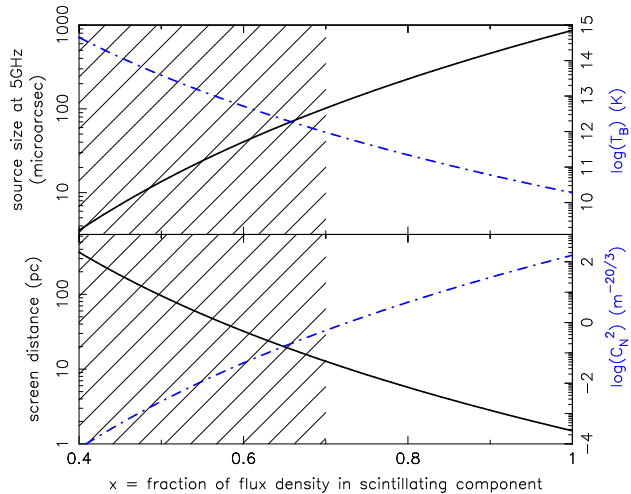}
\caption{The derived range of source and screen properties (upper and
  lower panel respectively), taking into account the effects of source resolution
  on the observed timescale and modulation index at the frequency of
  maximum variation. The solid lines refer to the scale on the left, the dotted lines to that on the right. We require that the combined effects of the
  non-scintillating component and the size of the scintillating
  component result in the observed maximum modulation index, and assume
  $\theta\propto\nu^{-0.5}$.  The values at $x=0.4$ correspond to a
  point-like ($\theta_S < \theta_{\rm scatt}$) component on a
  non-scintillating base. However, the left-hand side (shaded) is not
  compatible with other considerations of the data: the PDFs and the
  original lightcurves show this is not compatible with the deep
  minima: $\ge$ 70\% of the flux density must be in a scintillating
  component. }
\label{fig:limits}
\end{figure}

We have presented the boundary scenarios: the reality may lie
somewhere in between.  Fig.~\ref{fig:limits} shows the two extreme
cases discussed above ($x$=1 and 0.7), as well as the intermediary
cases. We note that the size quoted is the effective diameter, so for
an axial ratio of 6:1, the source would be $\sqrt 6 \approx 2.5$ times this
in the direction of the elongation, and $1/\sqrt 6 \approx 0.4$ times this
orthogonally to it.  

Our observed $\nu^{\rm o}$ at around 5\,GHz falls near to the critical
frequency predicted from contributions to scattering from the {\em
  entire} Galaxy \cite{wal98}, but this must be coincidental because,
aside from the reasons given in this section, the peculiar velocity
rules out a contributions from a very extended material, as does the
slope of the structure function. We calculate a high critical frequency
of between 13 and 25\,GHz ($0.7 < x <1$). In this particular direction
we see that any source brighter than a hundred mJy at 5\,GHz would
have to have a T$_B$ far in excess of the Compton limit if it were to
demonstrate an observed modulation maximum around 25\,GHz. The
coincidence is probably also partly a reflection of the more frequent
observing at GHz frequencies, where this source is most variable.

We consider also the possibility of the source as modelled by two
well-separated sources, each smaller than the scattering disk but
separated by several times this. Such a situation is similar to some
structures observed in VLBI images of core-jets and compact radio
galaxies. Motion parallel to the axis of separation of the components
would produce a conspicuous `dip' in the rising portion of the structure
function \citep{spa93}. The nature of the observed structure functions
for \source indicates this is unlikely, as the fit indicate most
observations as well as most scintles are observed during motion
suitable to see the development of these two timescales. There is no
such evidence for this: the 'overshoot' of the structure function
cannot be explained in these terms, as the structure function reaches
saturation values before turning over. We therefore rule out a model
of discrete and well-separated source components.

Finally we note that, as well as not using anisotropic scattering
theory, all these calculations are based on asymptotic theory which is
not strictly applicable in this regime.  Further work testing the
theory with simulations near the critical frequency is required, as
well as the effects on the observables of partially and slightly
resolved sources ($\theta_S \approx \theta_{\rm scatt}$).  Only more
careful theoretical work will enable the scattering material and
source size to be properly constrained. At present, even for a source
with very well defined characteristics (a strong limit on
`non-scintillating' components, and a known velocity of the scattering
screen) the constraints on source size are not tight.

It can be seen that the brightness temperature can be below the
Compton limit, but only if the scatterer is very strong and only a few
parsecs from Earth.  Regardless of the scattering theory, we note that
if the source has a T$_B <5\times 10^{11}$K, the source must be $>$
140$\mu$as in diameter. The detection of anisotropy, whether in
scatterer or source, is a robust result which does not depend on
theoretical details. The anisotropy required by the annual modulation
fits has an axial ratio of $>$ 6:1, and if this is due to source
structure, the longest axis must be $>$ 0.34\,mas at 5\,GHz. This is
resolvable by space VLBI or ground-based VLBI at higher frequencies.
Global VLBI observations at 22\,GHz (Dennett-Thorpe, Gurvits et al. in
prep) show \source to be $<$0.4\,mas at this frequency.  If we assume
that the source at 22\,GHz closely related to that at 5\,GHz, in
particular the (not unreasonable) assumption that there are no
milliarcsecond components which are present at 5\,GHz and not at
22\,GHz, we determine an upper limit on the source size at 5\,GHz.
Using the relation source size $\propto \nu^{-0.5}$, we find the major
axis of the source at 5\,GHz is $<$ 8\,mas, which is an equivalent
diameter for axial ratio 6:1 of $<$0.32\,mas. From this we favour a
source which exceeds $5\times10^{11}$K at 5\,GHz. Combined with the
preceeding discussion this would put the screen at a distance between
4 and 12\,pc.

\section{Concluding remarks}

We have presented observations over a two year period for the quasar
J1819+3845. Notable results of this study were:

(i) The source continues to vary throughout the two years.
Although the timescale of these variation changes by a factor of ten
or more, the strength of the modulation does not appear to change.

(ii) The change in timescale has a periodicity of a year, and can be
explained by a velocity of the plasma screen and an additional
component, which is due to source structure or anisotropic turbulence
in the scattering plasma. These observations do not allow us to
uniquely determine all four parameters, although v$_{\rm dec}$ of the
scattering plasma and the position angle of the anisotropy are well
constrained. Combining these observations with the results of our
previous experiment \citep[, Paper II]{den01} where we measured the
velocity of the plasma screen directly, we constrain the axial ratio
of the source, or the turbulent cells to be $>$ 6:1 and oriented at
PA=$83^\circ\pm4^{\circ}$.

(iii) The source is resolved, as can be seen by the smoothness of the
modulations. From the depth of the modulations, we deduce that the
source at 5\,GHz is at most a few times the scattering disk at the
scattering screen.

(iv) Based on the interpretation of the structure function, we
determine that the plasma responsible for scattering is a thin screen.

(v) Any source expansion has been $<$ 10\% over the two years of
monitoring. No significant change in structure has been detected.

(vi) The source has been getting brighter at 5\,GHz. The increase in
flux density has been $\sim$ 25\%/year. Coupled with the lack of
source expansion, this 50\% increase in flux density, must be largely due to an increase in the brightness temperature.

Interstellar scintillation has already been proved to be the cause of
the variations in J1819+3845 (Paper II), but this paper shows that
monitoring observations over a year can also be a proof of
scintillation. Since this project started this method has been applied
to B\,0917+624 \citep{ric01b,jau01}, although using a very few points.
Monitoring campaigns are more appropriate for the majority of IDV
sources, as their timescales of modulation are not short enough to be
able to conduct a two-telescope delay experiment. Furthermore, we
demonstrate that the transverse velocity of the scattering plasma can be
calculated using this method: to our knowledge this is the first such
measurement.

Using the determined plasma velocity we obtain an equivalent angular
source size of between 100 and 900\,microarcseconds, and corresponding
brightness temperatures in the range 10$^{10}$ and 10$^{12}$K. In this
calculation we consider the effects of the spatial distribution of the
scintillating flux -- how much of the source could be smaller than the
scattering disk.  This consideration is the main reason for a decrease
in T$_B$ from 10$^{14}$K calculated in Paper I, although we still
favour a source T$_B > 5\times10^{11}$K. The scattering material is
very unusual: it is very strong and between 1 and 12 parsecs away.

Modelling of the source extent in relation to the the scattering disk
is a critical factor in determining both the intrinsic source size and the
parameters of the scattering medium, and we are undertaking work on this.
Although we have taken account of source resolution, we have not used
a full and proper approach to deal with the anisotropy, and this also needs
to be addressed.

This work emphasises that consideration of the effects of
source extension could have a very big impact on the parameters
derived. In the meantime we must be wary of using models of
scintillating and non-scintillating components to describe
scintillating extragalactic radio sources, as the source sizes and
brightness temperatures obtained in such a way are very poor
estimates. 

The detection of the anisotropy is a robust result which does not
depend on the details of the theory, but whether the anisotropy is due
to the source itself, or to the scattering material remains an open
question.

\begin{acknowledgements}

This research was supported by the European Commission, TMR
Programme, Research Network Contract ERBFMRXCT96-0034 `CERES' and NOVA.  The
William Herschel Telescope is operated on the island of La Palma by
the Isaac Newton Group in the Spanish Observatorio del Roque de los
Muchachos of the Instituto de Astrofisica de Canarias. The WSRT is
operated by the Netherlands Foundation for Research in Astronomy
(NFRA/ ASTRON) with financial support by the Netherlands Organization
for Scientific Research (NWO). 

We wish to thank Frank Briggs for providing us with the solution to
the 3D trigonometry; Rene Vermeulen for generous scheduling at the
WSRT especially in the early stages of the project; and Barney
Rickett and J-P Marquart for helpful discussions. JDT also
thanks the Astronomical Institute Anton Pannekoek, University of
Amsterdam for their hospitality, and the Netherlands Research School
for Astronomy (NOVA) for support during the completion of this work.
 
\end{acknowledgements}

\end{document}